\documentclass[10pt,aps,amsfonts,pra,twocolumn,showpacs]{revtex4-1}
\usepackage{verbatim,epsfig,amsmath,amssymb,bm,epsf,graphicx,psfrag,bbold,amsthm,amsfonts}
\usepackage{hyperref}
\usepackage[all]{xy}
\usepackage{color}

\newcommand{\ket}[1]{|#1\rangle}

\newcommand{\innerproduct}[2]{\langle #1| #2\rangle}

\newcommand{\dagr}[1]{{#1}^{\dagger}}
\newcommand{\ztzt}{\mathbb{Z}_2 \times \mathbb{Z}_2}
\newcommand{\ztwo}{\mathbb{Z}_2}
\newcommand{\zn}{\mathbb{Z}_N}
\newcommand{\zthree}{\mathbb{Z}_3}
\newcommand{\onsitesymmetrygroup}{G_{int}}
\newcommand{\parity}{\mathcal{P}}
\newcommand{\paritygroup}{G_\mathcal{P}}
\newcommand{\translation}{\mathcal{L_T}}
\newcommand{\inversion}{\mathcal{I}}
\newcommand{\timerev}{\mathcal{T}}
\newcommand{\timerevgroup}{G_\mathcal{T}}
\newcommand{\conjug}{\mathcal{\theta}}

\newcommand{\af}{A_4}
\newcommand{\hgc}{H^2(\onsitesymmetrygroup,U(1))}
\newcommand{\hgcinput}[1]{H^2(#1,U(1))}

\newcommand{\thatis}{$i.e.~$}

\begin{document}

\title{Detection of gapped phases of a 1D spin chain with onsite and spatial symmetries}

\author{Abhishodh Prakash}
\affiliation{C. N. Yang Institute for Theoretical Physics and
Department of Physics and Astronomy, State University of New York at
Stony Brook, Stony Brook, NY 11794-3840, USA}

\author{Colin G. West}
\affiliation{C. N. Yang Institute for Theoretical Physics and
Department of Physics and Astronomy, State University of New York at
Stony Brook, Stony Brook, NY 11794-3840, USA}

\author{Tzu-Chieh Wei}
\affiliation{C. N. Yang Institute for Theoretical Physics and
	Department of Physics and Astronomy, State University of New York at
	Stony Brook, Stony Brook, NY 11794-3840, USA}

\date{\today}

\begin{abstract}
	We investigate the phase diagram of a quantum spin-1 chain whose Hamiltonian is invariant under a global onsite $A_4$, translation and lattice inversion symmetries. We detect different gapped phases characterized by SPT order and symmetry breaking using matrix product state order parameters. We observe a rich variety of phases of matter characterized by a combination of symmetry breaking and symmetry fractionalization and also the interplay between the onsite and spatial symmetries. Examples of continuous phase transitions directly between topologically nontrivial SPT phases are also observed.
\end{abstract}

 \maketitle
 
 \section{Introduction}
 The program of classifying and characterizing different phases of matter has been revived and actively pursued in recent years. One aspect is to classify phases based on global symmetries. In the Landau-Ginzburg paradigm, given a class of many-body Hamiltonians invariant under a global symmetry defined by a group $G$, different phases of matter can be enumerated by the \emph{spontaneous symmetry breaking} of $G$ and labeled by the residual symmetry $H$ that $G$ is broken down to. One could also envision the existence of local order parameters which arise from symmetry breaking and hence be able to distinguish between these phases. However, after the discovery of the Quantum Hall Effect~\cite{IQHE_Laughlin,FQHE_Laughlin}, it was realized that the Ginzburg-Landau symmetry-breaking picture might not be enough to classify all phases of matter~\cite{Wenbook}. Some systems like the fractional quantum hall states~\cite{FQHE_Laughlin}, spin liquids~\cite{Spin_Liquid_Kalmeyer_Laughlin}, quantum double models~\cite{Toric_code_Kitaev} and string-net models~\cite{String_net_Wen_Levin} do not even need symmetries and are called \emph{intrinsic topological phases} or simply \emph{topological phases}. Even with symmetries, several new phases have been discovered which are not classified by symmetry-breaking or characterized by local order parameters; such as topological insulators~\cite{Kane_Topins} and the Haldane phase of spin-1 chains~\cite{HaldanePhase1,HaldanePhase2,aklt} and these phases are called \emph{symmetry protected topological} (SPT) phases~\cite{wen_Tensor_Entanglement, pollman_turner_entspectrum, pollmann2012symmetry}.  Furthermore, if we consider global symmetry in systems with intrinsic topological order, we can have more phases called \emph{symmetry enriched topological phases}~\cite{set_mesaros,set_essin,set_stringflux}. In gapped 1D spin chains, which we focus on in this paper however, it has been shown that there cannot be any intrinsic topological order and hence all phases are either symmetry breaking or SPT phases~\cite{wen_Tensor_Entanglement, top1d_wenold,top1d_wennew,top1d_lukasz,top1d_pollmanturner,top1d_schuchgarciacirac}. 
 
 Given that the classification program has been much explored, there has been interest in developing ways to detect which phase of matter a system belongs to. Since local order parameters are insufficient to detect phases that are not characterized by spontaneous symmetry breaking (SSB), there have been attempts to develop other quantities that can detect SPT phases like non-local `string' order parameters~\cite{string_order_deNijs,string_order_kennedy1992,string_order_garcia,string_order_hagheman,top1d_pollmanturner} and Matrix Product State (MPS) order parameters~\cite{top1d_pollmanturner}. Furthermore, if we include the possibility of both symmetry breaking and SPT phases, there is a rich set of possible phases~\cite{top1d_wennew}. Given a global symmetry group $G$, the ground state can spontaneously break the symmetry to one of its subgroups $H \subset G$. However for each subgroup $H$, there can exist different SPT phases that do not break symmetry spontaneously. The situation is even more interesting if there are both internal and space-time symmetries like parity and time reversal invariance. In this paper, we generalize the techniques of Ref~\cite{top1d_pollmanturner} and study the phase diagram for a two parameter Hamiltonian of a spin-1 chain which is invariant under a global onsite (internal) $A_4$ symmetry, lattice translation and lattice inversion (parity). Through suitable order parameters, we detect both the different SSB and SPT phases and label them using the classification framework of Ref~\cite{top1d_wennew}. A total of eight distinct phases are identified within the parameter space we consider. In particular, we find among these a direct, continuous transition between two topologically nontrivial $A_4$-symmetric SPT phases, distinguished by the 1D representations of the symmetries, as explained below. 
 
 This paper is organized as follows. In section \ref{Overview}, we described the $A_4$ spin-chain Hamiltonian studied here and present its phase diagram which contains the main results of this paper. In section \ref{sec:Classification}, we review the classification of 1D gapped-spin chains and list parameters which can be used to completely classify phases. In section \ref{sec:FullResults}, we describe the full details of the phase diagram of the $A_4$ model, and also enumerate the several possible phases that can in principle exist given the symmetry group of the parent Hamiltonian. Section \ref{sec:Numerical} presents, in detail, the numerical techniques by which the states and parameters were obtained, and section \ref{Summary} gives a summary of our results.

 \section{Overview of main results}\label{Overview}
 
 \subsection{ The Hamiltonian}
 We will now describe an $\af$ and inversion symmetric Hamiltonian whose phase diagram we study in detail. The Hamiltonian we present here is a modified version of the one used in Ref~\cite{abhishodh1} where it was found that the there was an extended region where the ground state is exactly the AKLT state and hence useful for single qubit quantum information processing~\cite{miyake_aklt}. Here, we slightly modify the Hamiltonian to retain the essential features only and study the phase diagram. 
 
 The total Hamiltonian consists of three parts. The first is the Hamiltonian for the spin-1 Heisenberg antiferromagnet which is invariant under the spin-1 representation of $SO(3)$:
 \begin{eqnarray}
 H_{Heis} &=& \sum_i  \vec{S}_i\cdot\vec{S}_{i+1}  , \\
 \end{eqnarray}
where~$\vec{S}_i\cdot\vec{S}_{i+1} \equiv S^x_i S^x_{i+1} + S^y_i S^y_{i+1} + S^z_i S^z_{i+1}$. We add two other combinations, $H_q$ and $H_c$ to the Heisenberg Hamiltonian which breaks the $SO(3)$ symmetry to $\af$, the alternating group of degree four and the group of even permutations on four elements (equivalently, the rotation group of a tetrahedron). These terms are defined as:

 \begin{eqnarray}
 H_q &=& \sum_i  (S^x_i S^x_{i+1})^2 + (S^y_i S^y_{i+1})^2 + (S^z_i S^z_{i+1})^2,\nonumber
 \end{eqnarray}
 and
 \begin{multline}
 H_c = \sum_i [ (S^x S^y)_i S^z_{i+1} + (S^z S^x)_i S^y_{i+1} + (S^y S^z)_i S^x_{i+1} \\
 + (S^y S^x)_i S^z_{i+1} + (S^x S^z)_i S^y_{i+1} + (S^z S^y)_i S^x_{i+1} \\
 + S^x_{i} (S^y S^z)_{i+1} + S^z_{i} (S^x S^y)_{i+1} + S^y_{i} (S^z S^x)_{i+1}  \\
 +  S^x_{i} (S^z S^y)_{i+1} + S^z_{i} (S^y S^x)_{i+1} + S^y_{i} (S^x S^z)_{i+1}]. 
 \end{multline}
 
For details on how the perturbations are constructed, see Appendix~\ref{Hamiltonian_construction} or Ref~\cite{abhishodh1}.
 
 The operators in $H_c$ are symmetrized so that the Hamiltonian is invariant under inversion as well as lattice translation. With this we have a two-parameter Hamiltonian invariant under an onsite $A_4$ symmetry along with translation invariance and inversion.
 \begin{equation}
 \label{eq:H}
 H(\lambda, \mu) = H_{Heis} + \lambda H_c + \mu H_q.
 \end{equation}

 \subsection{Summary of numerical results}\label{Results}
 
\begin{figure}[ht]
	\includegraphics[width = 90mm]{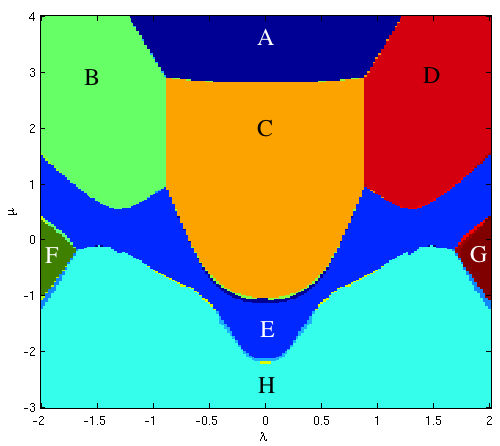}
	\caption{\label{fig:PhaseDiagram} (Color Online). The phase diagram for a two-parameter Hamiltonian constructed to have an $A_4$ onsite symmetry group, as well as parity symmetry and one-site translation invariance. The symmetries of the Hamiltonian break down into five different residual symmetry groups in the ground states. These break down further when classified according to the relevant topological parameters, yielding eight distinct phases overall. The diversity of phases from the comparatively simple Hamiltonian shows the necessity of carefully accounting for all possible symmetries and topological parameters when attempting to characterize the phase of a ground state. For a description of the phases \textbf{A-H}, see discussions in the main text.}
\end{figure}

 We employ the iTEBD algorithm\cite{iTEBD} to numerically analyze the  ground states across a range of parameters $\mu = [-3, 4]$, $\lambda = [-2, 2]$ and find a wide variety of phases. In the parameter space analyzed, a total of eight distinct regions can be identified (labeled with letters \textbf{A-H} in Fig.~\ref{fig:PhaseDiagram}). These regions are distinguished both by the symmetries of the ground states, and also by the classification parameters of Ref~\cite{top1d_wennew}.
 
 From the symmetry group $G$ of the parent Hamiltonian, which contains $A_4$, spatial inversion, and translation symmetries, only the inversion and translation symmetries remain in the ground states of region \textbf{A}. Regions \textbf{B}, \textbf{C}, and \textbf{D}, by contrast, all respect the full set of symmetries of the parent Hamiltonians but are differentiated by one of the SPT parameters: namely, the overall complex phase produced under $A_4$ transformations. These complex phases are different 1D irreducible representations (irreps) of $\af$ and correspond to distinct SPT phases protected by translation and onsite symmetries. In phase \textbf{E}, the ground state breaks the symmetry to onsite $\ztwo$ and parity. The translation symmetry in this region is broken down from single-site translation invariance to two-site. This broken, two-site translation symmetry is also present in regions \textbf{F} and \textbf{G}, but here the remaining symmetries of $\af$ and parity are completely preserved. Like regions \textbf{B}, \textbf{C}, and \textbf{D}, regions \textbf{F} and \textbf{G} have the same symmetry but are distinguished from one another only by the values of their SPT parameters. Finally, in region \textbf{H}, the residual symmetry group has an internal $\ztwo \times \ztwo$ symmetry and parity along with an one-site translation invariance.
 
 Among these eight phases, five correspond to instances of SSB and the remaining three correspond to SPT phases without symmetry breaking. The complete set of such parameters classifying these phases will be described in section \ref{sec:Classification}, and the particular values which distinguish them from one another are presented in section \ref{sec:FullResults}.
 
\begin{figure}[ht]
	\includegraphics[width = 90mm]{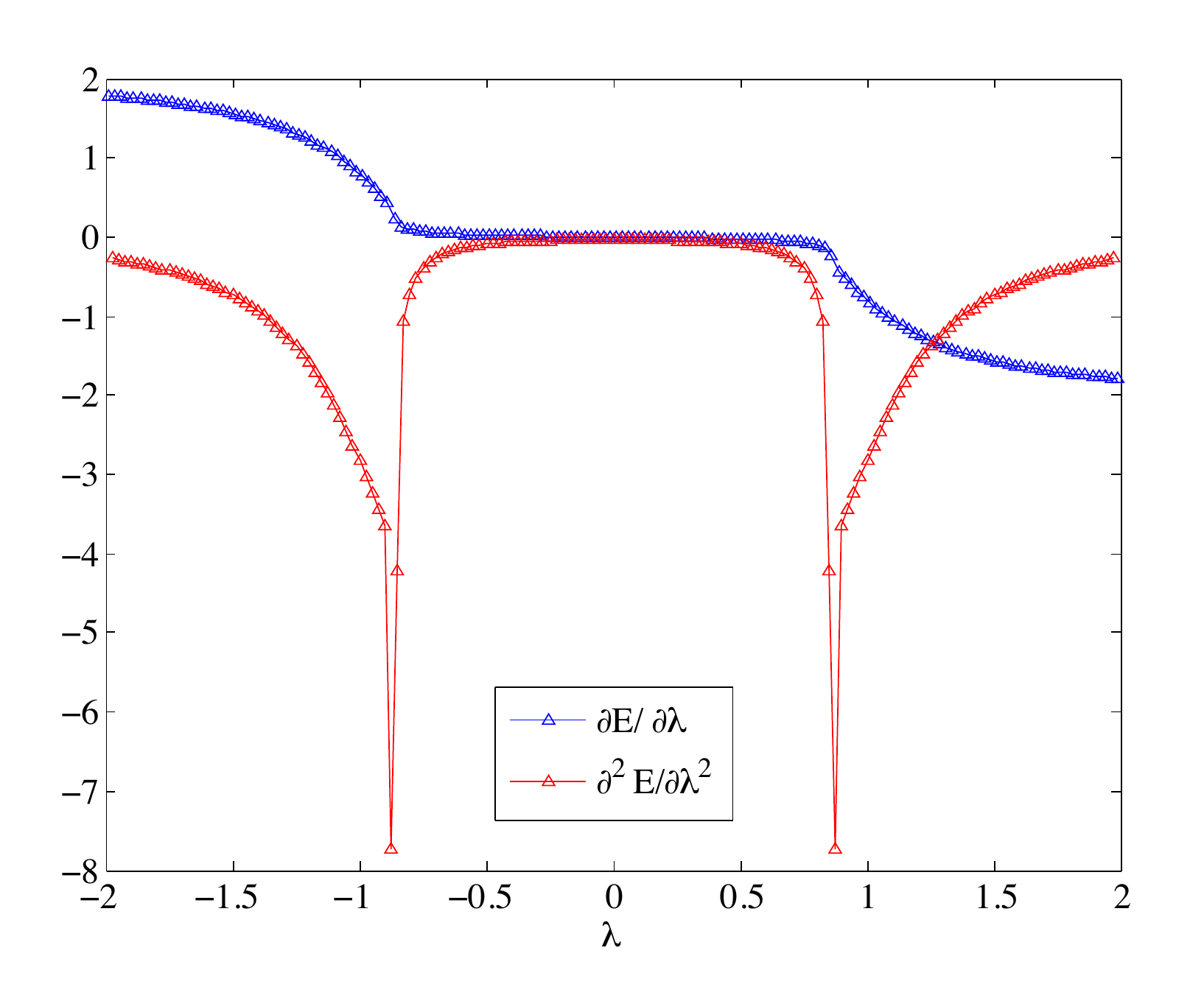}
	\caption{\label{fig:SecondOrder} (Color Online). Free energy derivatives along the line $\mu = 2$ in the phase diagram above show the nature of the phase transitions. The continuous first derivative (blue) contrasts with divergence in the second derivative (red), showing a second-order transition. All three regions are topologically nontrivial SPT phases. Data shown here was computed with a bond dimension of $30$, and the behavior has been seen to be stable as the bond dimension increases.}
\end{figure}

 Because the phases \textbf{B}, \textbf{C}, and \textbf{D} are not distinguished by any symmetry-breaking criteria (and because none of them are topologically trivial), the boundary lines between them are of particular interest as examples of non-trivial SPT to non-trivial SPT phase transitions. Such transitions are considered uncommon and have recently attracted particular interest\cite{SPTOTransChen, SPTOTrans2, SPTOTrans3,SPTOTrans4}, as compared to the more typical case of a transition between SPT and symmetry breaking phases, or trivial to non-trivial SPT phase transitions. Our analysis, however, shows that this model contains direct nontrivial SPT to SPT transitions, and that the transition is second-order in nature. By directly calculating the ground-state energy and its derivatives, we see sharp divergences in the second derivative, but a continuous first derivative across the boundary between these phases. Representative behavior is shown in Fig.~\ref{fig:SecondOrder}. 
 
\begin{figure}[ht]
 	\includegraphics[width = 90mm]{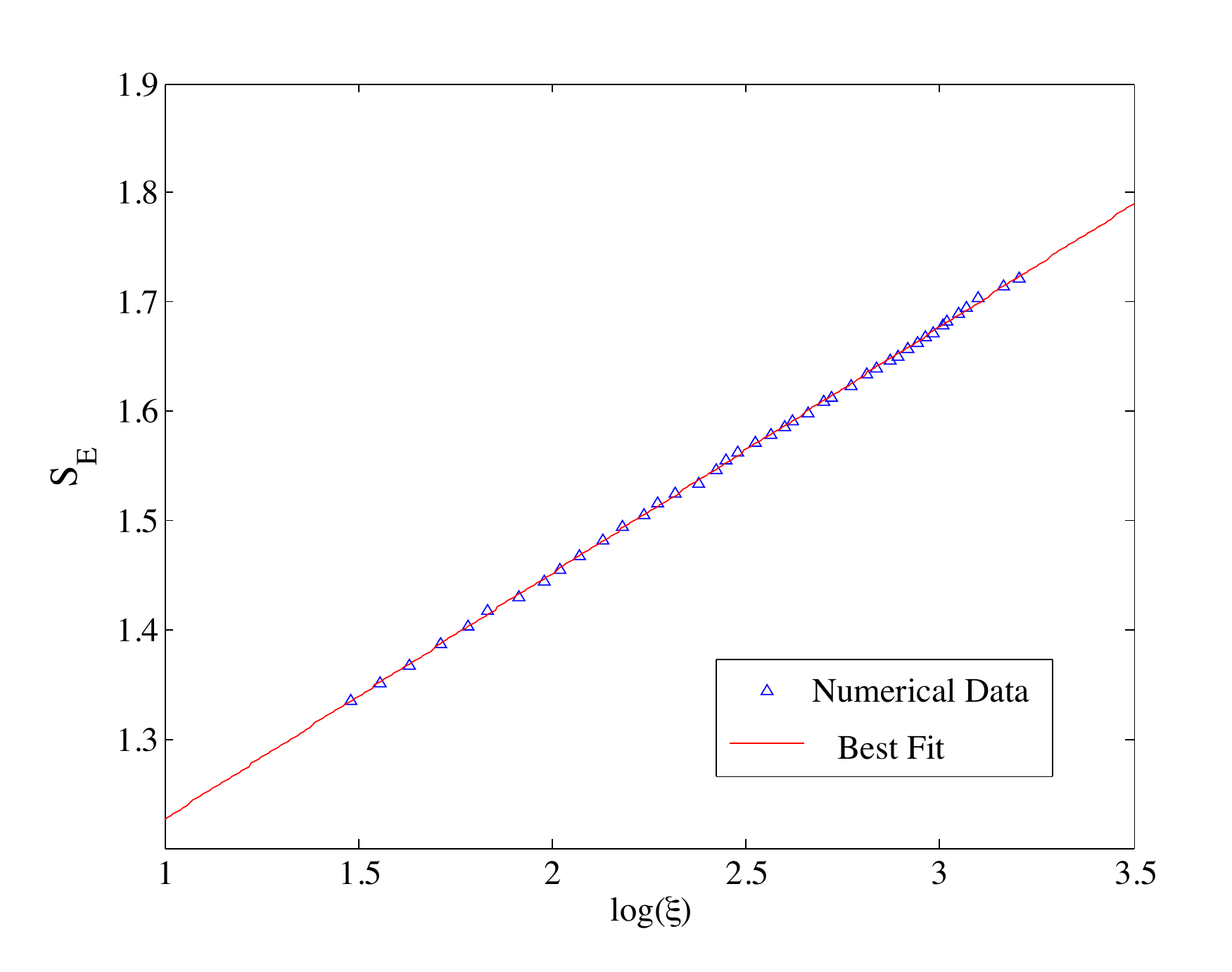}
 	\caption{\label{fig:Scaling} (Color Online). Entanglement entropy versus the log of the correlation length for states very close to the transition point. The slope is directly proportional to the central charge of the associated CFT, via Eq.~\ref{eq:ScalingRelation}. Data is generated by computing ground states at the point $\mu = 2, \lambda = 0.865$, and increasing the ``bond dimension" of the numerical scheme to allow us to find states closer to the critical point where the correlation length diverges. The behavior shown here is representative of that seen elsewhere along the lines $\lambda = \pm 0.865$. Away from these lines, the entanglement entropy saturates at a finite value of $\xi$. The best-fit line has a slope of $0.225(1)$, which corresponds to a central charge of $1.35(1)$. }
 \end{figure}

 The numerical methods employed here also allows us to probe the central charge of the conformal field theory (CFT) associated with the continuous phase transitions. As one approaches the transition, the correlation length begins to diverge. The central charge of the CFT appears in an important scaling relation between this diverging correlation length and the mid-bond entanglement entropy \cite{CalabreseCardy,pollmann-1dcritical}. In particular, it has been shown that
 \begin{equation} \label{eq:ScalingRelation}
 S =  \frac{c}{6}\log{\xi}
 \end{equation}
 where $c$ is the central charge, and $\xi$ is the correlation length measured in units of lattice spacing. $S$ is the entanglement entropy, given by  performing a Schmidt decomposition between sites and computing the entropy of the resulting Schmidt coefficients $\lambda_i$, 
 \begin{equation}
 S = -\sum_i \lambda_i \log{\lambda_i}.
 \end{equation}
 The MPS algorithms employed here to determine the ground state are not well-suited to computing ground states at the actual critical points. This is because the numerical accuracy of these algorithms are controlled by a tunable numerical parameter, the so called ``bond dimension." The closer we approach the critical point, the bigger this parameter needs be chosen for the ground states to be computed faithfully. By gradually increasing the bond dimension near the critical point, we obtain states with increasingly large correlation length, allowing us to fit the scaling relation of Eq.~\ref{eq:ScalingRelation}. We can also use this data to estimate the location of the transition, because away from the critical point, the scaling relation will not hold, and $S$ will saturate for large enough $\xi$ (or in practice, for large enough bond dimension). We find the critical lines to be located at $\lambda = \pm 0.865(2)$; fits at multiple points along these lines suggest a central charge of $c = 1.35(1)$, as shown for example in Fig.~\ref{fig:Scaling}.
 %Rephrase the above paragraph
 
 \section{Review of classification of 1D gapped phases of spin chains}\label{sec:Classification}
 We now review the classification of 1D gapped phases of spin chains following \cite{top1d_wennew}. Given the group of global symmetries $G$, the classification gives us a set of labels whose values distinguishes all possible phases of matter that can exist. We will systematically list these labels for various types of symmetries. It is the value of these labels that we extract numerically to determine the phase diagram presented in Sec~\ref{Overview}. First, we must give a brief introduction to \emph{matrix product state} (MPS) representations of one dimensional wavefunctions~\cite{MPS_Perez-Garcia2007}, which forms the backbone of the classification scheme. 
 \subsection{MPS formalism}
 Consider a one-dimensional chain of $N$ spins. If each spin is of $d$-levels \thatis the Hilbert space of each spin is $d$-dimensional, the Hilbert space of the spin chain itself is $d^N$-dimensional. A generic state vector in this many body Hilbert space is of the form
 \begin{equation} \label{eq:vector}
 \ket{\psi} = \sum_{i_1=1}^d \ldots \sum_{i_N=1}^d c_{i_1 \ldots i_N} \ket{i_1 \ldots i_n}
 \end{equation}
 This means that the number of coefficients $c_{i_1 \ldots i_N}$ needed to describe such a wavefunction grows exponentially with the length of the chain. To write this wavefunction in the MPS form, we need to associate for every spin site  (labeled by $m=1 \dots N$), a $D_m \times D_{m+1}$-dimensional matrix $A^{i_m}_m$ for each basis state $\ket{i_m} = \ket{1} \dots \ket{d}$ such that (assuming periodic boundary conditions without any loss of generality here and henceforth)
 \begin{eqnarray}\label{eq:coefficient_ansatz}
 c_{i_1 \ldots i_N} =  Tr[ A^{i_1}_1 A^{i_2}_2 \dots  A^{i_N}_N].
 \end{eqnarray}
 The matrices $A^{i_m}_m$ (which we will call MPS matrices) can, in principle always be obtained via sequential singular value decompositions of the coefficients $c_{i_1 \ldots i_N}$, as described in~\cite{vidal03}; in practice, it is useful to employ the canonical form of the MPS~\cite{MPS_Perez-Garcia2007, vidal03}. For most of the paper, the term ``MPS" shall refer to wavefunctions written in the form
 \begin{equation} \label{eq:mps_periodic}
 \ket{\psi} = \sum_{i_1 \dots i_N} Tr[ A^{i_1}_1 A^{i_2}_2 \dots  A^{i_N}_N] \ket{i_1 \ldots i_N} .
 \end{equation}
 
 Two important features of the MPS representation bear relevance to the numerical methods employed in this paper. The first is $D = \max_m(D_m)$, called the `virtual' or `bond' dimension, which in general may need to be very large. However, if the wavefunction is the ground-state of a gapped Hamiltonian and hence has a finite correlation length, it can be efficiently written as an MPS wavefunction whose bond dimension approaches a constant value that is independent of the size of the chain~\cite{vidal03,hastings2007area,arad_itai}. And as one approaches a critical point, where the correlation length diverges, an increasingly large bond dimension is required to faithfully capture the ground states. Even though the ground states at criticality therefore cannot be accurately represented by an MPS, one can employ the scaling results discussed above and in Fig.~\ref{fig:Scaling}, where increasingly large correlation lengths are probed by gradually increasing the bond dimension.
 
 Secondly, note that when a state possesses translation invariance, the MPS matrices themselves may be chosen to respect the same symmetry. A state invariant under one-site translations, for example, can be represented in the form above with the same MPS tensor at each site, $A_m^{i_m} = A^{i_m}$. This, in turn, allows a state with translation invariance of any length to be represented by $d$ matrices where $d$ is the dimension of the local Hilbert space. In general, a state with $K$-site translation invariance requires $Kd$ MPS matrices to represent it. 
 
 %There are times, however, when a larger number of matrices may be required, for example, when a state is being studied by a numerical algorithm which is not itself invariant on an $N$-site level. This, indeed, will frequently be the case for our states, as discussed in detail in sections \ref{sec:GroundStatePrep} and \ref{sec:OneSite}.

 \subsection{Symmetry breaking}
 First, we consider the possibility that the ground state spontaneously breaks the symmetry, $G$ of a Hamiltonian to $H$. This is the subgroup $H \subset G$ that still leaves the ground state invariant. This residual symmetry group itself acts as one of the labels to indicate the phase of matter. The case of the ground state not breaking any symmetry itself corresponds to $H=G$. However there may exist different SPT phases where the ground state is invariant under the same $H$. In such a case, we would need more labels along with $H$ to label the phase of matter. These labels depend on what $H$ itself is and will be reviewed next.
 
 We now consider the action of global symmetries on the physical spins and how it translates to the action on the MPS matrices on the virtual level. It was observed that the representation of the symmetry on the virtual level falls into distinct equivalence classes and these classes correspond to the different SPT phases of matter `protected' by the corresponding symmetry~\cite{pollman_turner_entspectrum,pollmann2012symmetry,top1d_wennew,top1d_lukasz,top1d_pollmanturner,top1d_schuchgarciacirac}. Here, we review the action of various symmetries on the MPS matrices, the different equivalence classes and the labels which distinguish them. The discussions here follow Ref~\cite{top1d_wennew}.
 \subsection{onsite/internal symmetry}\label{sec:onsite}
 Let us now consider Hamiltonians that are invariant under the action of a certain symmetry group $\onsitesymmetrygroup$ on each spin according to some unitary representation $u(g)$ \thatis $[H,U(g)]=0$ where $U(g) = u_1(g) \otimes  \cdots \otimes u_N(g) $. If the ground state $\ket{\psi}$ does not break the symmetry of the Hamiltonian, it is left invariant under the transformation $U(g)$ up to a complex phase
 \begin{equation} \label{eq:wavefunction_invariance}
 U(g) \ket{\psi} = \chi(g)^N \ket{\psi}.
 \end{equation}
 Eq.~(\ref{eq:wavefunction_invariance}) can be imposed as a condition on the MPS matrix level as~\cite{top1d_wenold, top1d_wennew, top1d_lukasz,top1d_pollmanturner}
 \begin{equation}
 \label{eq:mps_invariance}
 u(g)_{ij} A_M^j = \chi(g)V^{-1}(g) A_M^i V(g).
 \end{equation}
 Note that we use the Einstein summation convention wherein repeated indices are summed over. Because $u$ is a group representation, group properties constrain $\chi$ to be a 1D representation and $V$ generally to be a \emph{projective representation} of $\onsitesymmetrygroup$. A projective representation respects group multiplication up to an overall complex phase. 
 \begin{equation}\label{eq:projective_defn}
 V(g_1) V(g_2) = \omega(g_1,g_2) V(g_1 g_2).
 \end{equation}
 The complex phases $\omega(g_1,g_2)$ are constrained by associativity of group action and fall into classes labeled by the elements of the second cohomology group of $\onsitesymmetrygroup$ over $U(1)$ phases $\hgc$. In other words, \emph{the different elements of $\hgc$ label different classes of projective representations and hence different SPT phases of matter}. In particular, the identity element of the group $\hgc$ labels the set of \emph{linear representations} of $\onsitesymmetrygroup$ (which respect group multiplication exactly) and the corresponding phase of matter is trivial, containing or adiabatically connected to product ground states. 
 
 \subsection{Lattice translation}
 Note that we assume an infinite system with periodic boundary conditions for our discussions. 
 \subsubsection{Without onsite symmetry}
 The group of lattice translations $\translation$ is generated by single site shift $\mathcal{S}$ which acts as follows
 \begin{multline}
 \mathcal{S}: \sum_{i_1 \ldots i_N} c_{i_1 \ldots i_N} \ket{i_1 \ldots i_N} \rightarrow \sum_{i_1 \ldots i_N} c_{i_1 \ldots i_N} \ket{i_2, \ldots i_N, i_1} \\
 = \sum_{i_1 \ldots i_N} c_{i_N i_1 \ldots i_{N-1}} \ket{i_1 \ldots i_N}.
 \end{multline}
 In other words, 
 \begin{eqnarray}
 \mathcal{S} : c_{i_1 \ldots i_N} &\rightarrow&  c_{i_N i_1 \ldots i_{N-1}}\\
 \mathcal{S} : Tr[A_1^{i_1} \ldots A_N^{i_N}] &\rightarrow& Tr[A_2^{i_1}\ldots A_N^{i_{N-1}} A_1^{i_N}]
 \end{eqnarray}
 On the MPS matrix level, the single site shift acts as:
 \begin{equation}
 \mathcal{S} : A^i_M \rightarrow A^i_{M+1}.
 \end{equation}
 The full group, $\translation$ generated by $\mathcal{S}$ is
 \begin{eqnarray}
 \translation &=& \langle\mathcal{S} \rangle=\{e, \mathcal{S}^{\pm 1}, \mathcal{S}^{\pm 2}, \mathcal{S}^{\pm 3} \ldots \},\\
 S^{k} &:&  A^i_M \rightarrow A^i_{M+k}, ~~k\in\mathbb{Z}.
 \end{eqnarray}
 For a finite chain with periodic boundary conditions, we have the constraint $\mathcal{S}^N = e$ and hence $\translation \cong \zn$. For an infinite chain, $\translation \cong \mathbb{Z}$. It was shown~\cite{top1d_wennew} that \emph{there is only 1 SPT phase protected by $\translation$ alone. }
 
 If lattice translation is a symmetry, we can choose 
 \begin{equation}
 A^i_M = A^i_{M'} = A^i~~ \forall~ M,M' \in \{1,\ldots,N\},
 \end{equation}
 that is, the MPS matrices can be chosen to be independent of the site label and the same for all sites. 
 
 %Note however, as remarked above, that the matrices may not be chosen this way for numerical reasons; as described in Sec. \ref{sec:GroundStatePrep}, with the traditional iTEBD algorithm it is necessary to use a set of two-site invariant tensors even for one-site invariant states. Such two-site representations , however, can always be subsequently transformed into one-site representations through the block-diagonalization technique of Ref. \cite{MPS_Perez-Garcia2007}. Our approach to this is detailed in Sec. \ref{sec:OneSite}.
 
 \subsubsection{With onsite symmetry}
 If lattice translation is a symmetry in addition to onsite symmetry defined by a group $\onsitesymmetrygroup$ as described in Sec~(\ref{sec:onsite}), the different 1D irreducible representations (irreps) $\chi$ that can appear in Eq~(\ref{eq:mps_invariance}) also label different phases of matter. \emph{The different SPT phases protected by $G = \onsitesymmetrygroup\times \translation$ are labeled by $\{\omega, \chi\}$}
 
 \subsection{Parity}\label{sec:parity}
 \subsubsection{Without onsite symmetry}
 The action of inversion or parity, $\parity$ in general is generated by a combination of an onsite action by some unitary operator $w$ and a reflection, $\inversion$ that exchanges lattice sites about a point. 
 \begin{equation}\label{eq:parity_generator}
 \parity = w_1 \otimes w_2  \cdots \otimes w_N~\inversion
 \end{equation}
 where, the action of $\inversion$ is as follows:
 \begin{multline}
 \inversion: \sum_{i_1 \ldots i_N} c_{i_1 \ldots i_N} \ket{i_1 \ldots i_N} \rightarrow \sum_{i_1 \ldots i_N} c_{i_1 \ldots i_N} \ket{i_N i_{N-1}\ldots i_1}\\
 =  \sum_{i_1 \ldots i_N} c_{i_N \ldots i_1} \ket{i_1 \ldots i_N}.
 \end{multline}
 In other words,
 \begin{eqnarray}
 \inversion: c_{i_1 \ldots i_N} &\rightarrow& c_{i_N \ldots i_1}, \nonumber \\
 \inversion : Tr[A_1^{i_1} A_2^{i_2} \ldots A_N^{i_N}] &\rightarrow& Tr[A_1^{i_N} A_2^{i_{N-1}} \ldots A_N^{i_1}] \nonumber\\
 = Tr[&(A_N^{i_1})^T& (A_{N-1}^{i_2})^T \ldots (A_1^{i_N})^T]. \nonumber
 \end{eqnarray}
 In the last equation, we have used the fact that the trace of a matrix is invariant under transposition. On the MPS matrix level, the action is
 \begin{eqnarray}
 \inversion: A^i_M \rightarrow (A^i_{N-M+1})^T
 \end{eqnarray}
 The full action of parity is 
 \begin{eqnarray}
 \parity: c_{i_1 \ldots i_N} &\rightarrow& w_{i_1 j_1} \ldots w_{i_N j_N}  c_{j_N \ldots j_1}, \nonumber \\
 \parity: A^i_M &\rightarrow& w_{ij} (A^j_{N-M+1})^T 
 \end{eqnarray}
 Since $\parity^2 = e$, $w$ is some representation of $\ztwo$. There is a special lattice site that has been chosen as the origin about which we invert the lattice. It is sensible for parity to be defined without any reference to such a special point. Hence we assume that any system invariant under parity also has lattice translation invariance which allows any site to be chosen as the origin. Note that the action of inversion $\inversion$ and the generator of translations $\mathcal{S}$ do not commute. They are related by
 \begin{eqnarray}
 \inversion \mathcal{S} \inversion = \mathcal{S}^{-1}
 \end{eqnarray}
 The full symmetry group including translation invariance and parity, which we will call $\paritygroup$, generated by $\mathcal{S}$ and $\parity$ is (for a finite chain with periodic boundary conditions)
 \begin{multline}
 \paritygroup = \langle \parity, \mathcal{S} | \parity^2 = \mathcal{S}^N = e, \inversion \mathcal{S} \inversion = \mathcal{S}^{-1} \rangle \\ \cong \zn \rtimes \ztwo \cong D_N.
 \end{multline}
 For an infinite chain which we are interested in, we have
 \begin{multline}
 \paritygroup = \langle \parity, \mathcal{S} | \parity^2 = e, \inversion \mathcal{S} \inversion = \mathcal{S}^{-1} \rangle \cong \mathbb{Z} \rtimes \ztwo \cong D_\infty.
 \end{multline}
 If $\paritygroup$ is a symmetry of the Hamiltonian which is not broken by the ground state wavefunction $\ket{\psi}$, we have, under the action of $\parity$,
 \begin{equation}\label{eq:wavefuction_parity}
 \parity \ket{\psi} = \alpha(P)^N \ket{\psi}.
 \end{equation}
 The condition Eq.~(\ref{eq:wavefuction_parity}) can also be imposed on the level of the MPS matrices that describe $\ket{\psi}$:
 \begin{equation}\label{eq:mps_parity}
 w_{i j } (A^{j})^T = \alpha(P) N^{-1} A^i N,
 \end{equation}
 where, $\alpha(P) = \pm 1$ labels even and odd parity and $N$ has the property $N^T = \beta(P) N = \pm N$. \emph{$\{\alpha(P), \beta(P)\}$ label the 4 distinct SPT phases protected by $\paritygroup$}~\cite{top1d_wennew}.
 
 \subsubsection{With onsite symmetry}\label{sec:onsite+parity}
 Let us consider invariance under the combination of an onsite symmetry $\onsitesymmetrygroup$ as described in Sec(\ref{sec:onsite}) with parity. If the actions of the two symmetry transformations commute on the physical level,
 \begin{equation}
 U(g) \parity \ket{\psi} = \parity U(g) \ket{\psi},
 \end{equation}
 \thatis $G = \onsitesymmetrygroup \times \paritygroup$, this imposes constraints on the matrix $N$ defined in Sec.~\ref{sec:parity} as~\cite{top1d_wennew}.
 \begin{equation}\label{eq:onsite+parity_MPS}
 N^{-1} V(g) N = \gamma_P(g) V^*(g).
 \end{equation}
 Where, $\gamma_P(g)$ is a one-dimensional irrep of $\onsitesymmetrygroup$ that arises from the commutation of onsite and parity transformations~\cite{top1d_wennew} and $V(g)$ is the representation of $\onsitesymmetrygroup$ acting on the virtual space as discussed in Sec~\ref{sec:onsite}. Note that we can rephase $V(g) \mapsto \alpha(g) V(g)$ without changing anything at the physical level. However, Eq~\eqref{eq:onsite+parity_MPS} is modified replacing $\gamma_P(g) \mapsto \gamma_P(g)/\alpha^2(g)$. Hence, the 1D irreps $\gamma_P$ and $\gamma_P/\alpha^2$ are equivalent labels for the same phase for all the 1D irreps  $\alpha$ of $\onsitesymmetrygroup$. \emph{Different SPT phases of matter protected by $G = \onsitesymmetrygroup\times \paritygroup$ are labeled by $\{\omega,\chi(g),\alpha(P),\beta(P),\gamma_P(g) \}$}~\cite{top1d_wennew}. Where, as defined before $\omega \in \hgc $ with $\omega^2 = e$ and $\gamma_P \in \mathcal{G}/\mathcal{G}^2$ where $\mathcal{G}$ is the set of 1D representations of $\onsitesymmetrygroup$.
 
 Because our Hamiltoinian is not invariant under time reversal, we do not review the classification of SPT phases protected by time-reversal invariance and combinations with other symmetries here. We include the same in the Appendix~(\ref{app:timerev}) for the sake of completion. The techniques used in this paper can be extended easily to include time-reversal invariance. 
 
 \section{Using the parameters to understand the phases of the $A_4$ Hamiltonian}\label{sec:FullResults}
 \subsection{Details of the phase diagram}
 Armed with the family of parameters described in the last section, $\{\omega, \chi, \alpha, \beta, \gamma_P \}$, we now describe in detail the different phases of the Hamiltonian of Eq~(\ref{eq:H}) seen in Fig.~\ref{fig:PhaseDiagram}. The internal symmetry is $A_4$ which is a group of order 12 and can be enumerated by two generators with the presentation
 \begin{eqnarray}
 \innerproduct{a,x}{a^3 = x^2 = (ax)^3 = e }.
 \end{eqnarray}
 The 3D representation of these generators are
 \begin{eqnarray}
 a = \begin{pmatrix}
 0 & 1 & 0\\
 0 & 0 & 1\\
 1 & 0 & 0
 \end{pmatrix}, ~ x = \begin{pmatrix}
 1 & 0 & 0 \\
 0 & -1 & 0\\
 0 & 0 & -1
 \end{pmatrix}
 \end{eqnarray}
 This can be visualized as the rotational symmetry group of the tetrahedron.
 
 First we briefly outline the steps followed:
 
 \begin{enumerate}
 	\item For every point in parameter space $\mu = [-3, 4]$, $\lambda = [-2, 2]$ of the Hamiltonian of Eq~(\ref{eq:H}), we use the iTEBD algorithm~\cite{iTEBD} to compute the ground state. 
 	\item We determine the residual symmetry $H \subset G$ of the full symmetry group $G = \af \times \paritygroup$ that leaves the ground state invariant. This includes checking the level of translation invariance, which may be broken down from one-site to two-site or beyond.
 	\item We determine the labels (subset of $\{\omega, \chi(g) \alpha(P), \beta(P), \gamma_P(g)\}$) that characterizes the fractionalization of residual symmetry and measure their values using the appropriate MPS order parameters.
 \end{enumerate}
 
 Several of these steps involve important numerical considerations. Full details of our implementation of these steps can be found in Sec~\ref{sec:Numerical}.
 
 We find that there are eight different phases in total. These phases, labeled ``\textbf{A}" through ``\textbf{H}" as indicated to match the phase diagram in Fig.~\ref{fig:PhaseDiagram}, are characterized as follows:
 
 \begin{enumerate}
 	\item Phase \textbf{A:} Parity and one-site translation only \thatis $H= \paritygroup$ (all internal symmetries are broken). This region is therefore classified by the values of $\{\alpha(P), \beta(P) \}$ and is found to have values
 	\begin{itemize}
 		\item $\{ \alpha(P) = -1, \beta(P) = -1 \}$
 	\end{itemize}
 	\item \textbf{Phases B, C, and D:} No unbroken symmetries. The ground state in these three regions are invariant under the full symmetry group $G = \af \times \paritygroup$. The relevant labels are $\{\omega, \chi(g), \alpha(P), \beta(P)\}$ (Since all three 1D irreps of $A_4$ are equivalent under the relation $\gamma_P \sim \gamma_P/\chi^2$, $\gamma_P(g)$ is a trivial parameter). The MPS matrices in all three regions transform projectively \thatis these are non-trivial SPT phases with $\omega = -1$ where $\hgcinput{A_4} \cong \ztwo \cong \{1,-1\}$. Also, $\alpha(P)=-1, \beta(P)=-1$ for all three phases. However, they can be distinguished by the values of  $\chi$, \thatis observing that the 1D irrep produced under the $A_4$ symmetry transformation (Eq~(\ref{eq:wavefunction_invariance})) in the three regions corresponds to the three different 1D irreps of $A_4$ .  The values of the set of parameters which characterize the regions are as follows.
 	\begin{itemize}
 		\item Phase \textbf{B}: 
 		$\{\omega = -1, \chi: \{ a= e^{ \frac{i 2 \pi}{3}}, x=1 \}, \alpha = -1 , \beta = -1\}$
 		\item Phase \textbf{C}: 
 		$\{\omega = -1, \chi: \{a=1, x=1 \}  , \alpha = -1 , \beta = -1 \}$
 		\item Phase \textbf{D}:
 		$\{\omega = -1, \chi: \{ a= e^{- \frac{i 2 \pi}{3}}, x=1 \}, \alpha = -1 , \beta = -1\}$
 	\end{itemize}
 	
 	\item \textbf{Phase E:} Parity, $\ztwo$ and two-site translation. This region possess a hybrid parity $\paritygroup$, generated not by inversion alone but rather the combination of inversion and the order 2 element $axa^2$ of $\af$. Additionally, there is an unbroken onsite $\ztwo$ actions with elements $\{e, x\}$. The relevant labels are $\{\chi(g), \alpha(P), \beta(P), \gamma_P(g)\}$ with values 
 	\begin{itemize}
 		\item $\{\chi : \{e =1, x=1 \} , \alpha = 1 , \beta = 1, \gamma_P = \{e=1, x=1 \} \}$
 	\end{itemize}

 	\item \textbf{Phases F and G:} These regions possess the same parity and onsite $A_4$ symmetry as phases \textbf{B}, \textbf{C}, and \textbf{D}, but have translation invariance which is broken down to the two-site level. They are also distinct from the above phases because the MPS matrices transform under a linear representaion of $A_4$, and have a trivial representation of parity at the two site level. The relevant labels are parameters are $\{\omega, \chi(g), \alpha(P), \beta(P)\}$ with values
 	\begin{itemize}
 		\item Phase \textbf{F}: $\{\omega = +1, \chi: \{a= e^{- \frac{i 2 \pi}{3}}, x=1 \}, \alpha = +1, \beta = +1  \}$
 		\item Phase \textbf{G}: $\{\omega = +1, \chi: \{a= e^{+ \frac{i 2 \pi}{3}}, x=1 \}, \alpha = +1, \beta =  +1\}$
 	\end{itemize}
 	\item \textbf{Phase H:} In this final region, the onsite symmetry is broken down to a $\ztzt$ subgroup with elements $\{e, x, a^2 x a , axa^2\}$. Parity and translation symmetry are both fully retained. It is therefore the only region in our sample phase diagram which requires all five labels $\{\omega, \chi, \alpha, \beta, \gamma_P \}$ to characterize. The values here are 
 	\begin{itemize}
 		\item 	$\{\omega = +1, \chi =\{1,-1,1,-1 \} , \alpha = +1 , \beta = +1 , \gamma_P = \{1,1,1,1 \} \}$
 	\end{itemize}
 \end{enumerate}
 Note here that for compactness, the set of values given $\chi$ and $\gamma$ refer to the four elements $\{e, x, a^2 x a , axa^2\}$, respectively.
 
 The diversity of phases seen in this phase diagram show the importance of carefully checking for both conventional symmetry-breaking phases and SPT phases. The phases present here also underscore the importance of considering the different possible instances of parity and translation invariance which can occur, since in addition to traditional one-site translation invariance and inversion, one might find e.g. translation breaking without inversion breaking (phases \textbf{F} and \textbf{G}), or inversion which only exists when hybridized with an onsite symmetry (phase \textbf{E}). In the subsequent section, we show the wide variety of phases which could potentially exist given the symmetries of this parent Hamiltonian.
 
 \subsection{Counting the possible phases of with onsite $\af$ and parity symmetries}\label{sec:Counting}
 Looking beyond the eight phases which are observed in our phase diagram, we now count the different phases that are possible with the symmetry group that we considered. 
 \begin{eqnarray}
 G = \af \times \paritygroup
 \end{eqnarray} 
 $\paritygroup$ is the group generated by lattice inversion and translation that was described in Sec~(\ref{sec:parity}) and $\af$ is the alternating group of degree four \thatis the group of even permutations on four elements. The order of this group is 12 and can be enumerated with two generators, 
 \begin{eqnarray}
 \innerproduct{a,x}{a^3 = x^2 = (ax)^3 = e }.
 \end{eqnarray}
 We list the full set of 12 elements in terms of these generators for convenience: 
 \begin{multline}
 \af =\{e,a,a^2,ax,a^2x,xa,xa^2,xax,xa^2x,x,axa^2,a^2xa\} 
 \end{multline}
 
 We use the results of Refs.~\cite{top1d_wenold,top1d_wennew} which were reviewed in Sec~(\ref{sec:Classification}) and list the possible phases by the two possible mechanisms:
 \begin{enumerate}
 	\item Symmetry breaking of $G$ into the different possible subgroups.
 	\item SPT phases of the residual symmetry, $H$.
 \end{enumerate}
 
 The possible symmetry breaking patterns are enumerated by listing all possible subgroups of $G= \af \times \paritygroup$. Since in the thermodynamic limit, this is a group of infinite order formally isomorphic to $\af \times D_\infty$, we cannot list all possible subgroups. In particular, since the symmetry group we are interested in contains lattice translation $\translation$, generated by single site shifts, in principle, it can spontaneously break into various subgroups generated by two site shifts, three site shifts and so on when the ground state dimerizes, trimerizes etc. To keep things simple, here, we will list the possible phases where translation invariance is not broken. This means the effective group is $\af \times \parity$ which is formally isomorphic to $\af \times \ztwo$. Including all isomorphisms, there are 26 different subgroups contained in $\af \times \parity$ which can label the different residual symmetries, $H$ of the ground state at different levels of symmetry breaking. We shall list some of the non-trivial groups and the associated SPT phases. The counting would be similar when the ground state is invariant under, say, two-site translation invariance by considering two sites as one supersite and repeating the analysis similarly for the rest of the unbroken symmetry transformations. 
 \begin{enumerate}
 	\item\label{item:afour} $H = \af \times \parity$ :
 	\begin{itemize}
 		\item The set of labels that classify the different phases are $\{\omega, \chi, \alpha(P), \beta(P), \gamma_P \}$. 
 		\item For onsite $\af$, we have $\hgc = \ztwo$ which gives us two choices for $\omega$.
 		\item Since $\af$ has 3 different 1D irreps, we have three choices for $\chi$.
 		\item From the equivalence of 1D irreps described in Sec~(\ref{sec:parity}), all 1D irreps of $\af$ are equivalent to each other and $\gamma_P$ has only one choice.
 		\item $\alpha(P)$ and $\beta(P)$ have two choices each given by $\pm1$.
 		\item For $H = \af \times \parity$, we have $2 \times 3 \times 2 \times 2 \times 1 =$ \textbf{ 24 possible phases}.
 	\end{itemize}
 	\item \label{item:z2z2} $H = \ztzt \times \parity$ 
 	\begin{itemize}
 		\item There is one instance of $\ztzt \subset \af$ with $\ztzt = \{e, x, axa^2, a^2xa\}$.
 		\item The set of labels that classify the different phases are $\{\omega, \chi, \alpha(P), \beta(P), \gamma_P \}$. 
 		\item For onsite $\ztzt$, we have $\hgc = \ztwo$ which gives us two choices for $\omega$.
 		\item Since $\ztzt$ has 4 different 1D irreps, we have four choices for $\chi$.
 		\item Since each 1D irrep of $\ztzt$ squares to 1, all 4 of them are valid choices for $\gamma_P$.
 		\item $\alpha(P)$ and $\beta(P)$ have two choices each given by $\pm1$.
 		\item For $H = \ztzt \times \parity$, we have $2 \times 4 \times 2 \times 2 \times 4 =$ \textbf{ 128 possible phases}.
 	\end{itemize} 
 	\item \label{item:z3} $H = \zthree \times \parity$ 
 	\begin{itemize}
 		\item There are 4 instances of $\zthree \subset \af$: 
 		\begin{enumerate}
 			\item $\zthree^A = \{e, a, a^2\} $ 
 			\item $\zthree^B = \{e, xax, xa^2x\} $ 
 			\item $\zthree^C = \{e, ax, xa^2\} $ 
 			\item $\zthree^D = \{e, xa, a^2x\} $ 
 		\end{enumerate} 
 		\item The set of labels that classify the different phases for each instance are $\{\omega, \chi, \alpha(P), \beta(P), \gamma_P \}$. 
 		\item For onsite $\zthree$, we have $\hgc = \{e\}$ which gives us one choice for $\omega$.
 		\item Since $\zthree$ has 3 different 1D irreps, we have three choices for $\chi$.
 		\item From the equivalence of 1D irreps described in Sec~(\ref{sec:parity}), all 1D irreps of $\zthree$ are equivalent to each other and $\gamma_P$ has only one choice.
 		\item $\alpha(P)$ and $\beta(P)$ have two choices each given by $\pm1$.
 		\item For $H = \zthree$, we have $1 \times 3 \times 2 \times 2 \times 1 = 12$ phases for each of the four instances and hence a total of \textbf{ 48 possible phases}.
 	\end{itemize}   
 	\item \label{item:z2} $H = \ztwo \times \parity$ 
 	\begin{itemize}
 		\item There are 3 instances of $\ztwo \subset \af$: 
 		\begin{enumerate}
 			\item $\ztwo^A = \{e, x\} $ 
 			\item $\ztwo^B = \{e, axa^2\} $ 
 			\item $\ztwo^C = \{e, a^2xa\} $ 
 		\end{enumerate} 
 		\item The set of labels that classify the different phases for each instance are $\{\omega, \chi, \alpha(P), \beta(P), \gamma_P \}$. 
 		\item For onsite $\ztwo$, we have $\hgc = \{e\}$ which gives us one choice for $\omega$.
 		\item Since $\ztwo$ has 2 different 1D irreps, we have two choices for $\chi$.
 		\item Since each 1D irrep of $\ztwo$ squares to 1, both 1D irreps are valid choices for $\gamma_P$.
 		\item $\alpha(P)$ and $\beta(P)$ have two choices each given by $\pm1$.
 		\item For $H = \ztwo$, we have $1 \times 2 \times 2 \times 2 \times 2 = 16$ phases for each of the three instances and hence a total of \textbf{ 48 possible phases}.
 	\end{itemize}   
 	\item $H=$ Parity generated by lattice inversion only.
 	\begin{itemize}
 		\item There is one instance of this $H = \parity = \{e, \inversion\}$ \thatis all generators are $\af$ are broken.
 		\item The set of labels that classify the different phases are $\{\alpha(P), \beta(P)\}$. 
 		\item $\alpha(P)$ and $\beta(P)$ have two choices each given by $\pm1$.
 		\item For $H = \parity$, we have a total of $2 \times 2 =$ \textbf{4 possible phases}.
 	\end{itemize}
 	\item $H=$ Parity generated by lattice inversion combined with onsite $\ztwo$ operation.
 	\begin{itemize}
 		\item There are three possibilities: 
 		\begin{enumerate}
 			\item $\parity^A = \{e, x \inversion\}$
 			\item $\parity^B = \{e, axa^2 \inversion\}$
 			\item $\parity^C = \{e, a^2xa \inversion\}$
 		\end{enumerate}
 		\item The set of labels that classify the different phases for each instance are $\{\alpha(P), \beta(P)\}$. 
 		\item $\alpha(P)$ and $\beta(P)$ have two choices each given by $\pm1$.
 		\item For $H = \parity^{A/B/C}$, we have of $2 \times 2 = 4$ for each of the three instances and hence a total of \textbf{12 possible phases}.
 	\end{itemize}
 	\item $H = \ztwo \times \parity^{A/B/C}$
 	\begin{itemize}
 		\item There are three possibilities:
 		\begin{enumerate}
 			\item $\ztwo^A \times \parity^B = \{e, x, a^2 x a \inversion, a x a^2 \inversion\} $
 			\item $\ztwo^B \times \parity^C = \{e, x \inversion, a^2 x a \inversion, a x a^2 \} $
 			\item $\ztwo^C \times \parity^A = \{e, x \inversion, a^2 x a , a x a^2 \inversion\} $
 		\end{enumerate}
 		\item The set of labels that classify the different phases for each instance are $\{\omega, \chi, \alpha(P), \beta(P), \gamma_P \}$. 
 		\item With the number of choices for the labels being the same as mentioned in~(\ref{item:z2}), there are $1 \times 2 \times 2 \times 2 \times 2 = 16$ phases for each of the three instances and hence a total of \textbf{ 48 possible phases}.
 	\end{itemize}
 \end{enumerate}
 
 There are also other possibilities which include the different subgroups of $\af$ with parity being broken completely, in which case we ignore the labels $\alpha(P), \beta(P), \gamma_P$. It is clear that even in the limited set of residual symmetries we have listed, there is a rich set of phases when combined with SPT order.

\section{Numerical methods for obtaining the phase diagram}\label{sec:Numerical}

 For gapped 1D spin chains, the authors of Ref.~\cite{top1d_pollmanturner, pollman_turner_entspectrum, sukhi_injectivity,wen_Tensor_Entanglement} describe ways of numerically determining the SPT parameter  described above, and distinguishing different SPT orders. We build on the technique developed in Ref.~\cite{top1d_pollmanturner} where the authors obtain the SPT labels using the representations of symmetry at the virtual level. The numerical characterization of the phase diagram of a general parametrized Hamiltonian $H(\lambda, \mu, \ldots)$ proceeds according to the following steps:

\begin{enumerate}
	\item Identify the group of symmetries of the Hamiltonian, $G$ of the Hamiltonian.
	\item For each point in parameter space $\{ \lambda, \mu, \ldots\}$, obtain the ground state $\ket{\psi(\lambda,\mu, \ldots)}$ of the Hamiltonian $H(\lambda,\mu, \ldots)$ numerically as a MPS. 
	\item For each point in parameter space $\{ \lambda, \mu, \ldots\}$, identify the subgroup of symmetries $H \subset G$ that leaves the ground state $\ket{\psi(\lambda,\mu, \ldots)}$ invariant. In our case, this means checking each of the 24 elements of $G = A_4 \times \parity$. We also must explicitly check the translation invariance.
	\item Obtain the relevant virtual representations for the elements of $H$, \thatis $\chi, V, \alpha(P),$ and $N$. 
	\item From the representations and their commutation relations, obtain all other labels that completely characterize the phase. 
\end{enumerate}

In general, this process results in calculating the full family $\{\chi, \omega, \alpha(P), \beta(P), \gamma(P) \}$ for each point in parameter space. However, in some cases, the elements of $H$ are such that not all such parameters are necessary or even well-defined. For example, if the subgroup $H$ does not contain the parity operator, then $\alpha(P), \beta(P)$ and $\gamma(P)$ do not exist. Similarly, if $H = \zthree$, there is only one possible value of $\omega$, and hence we do not need it to distinguish the phase. The complete set of cases potentially relevant to our Hamiltonian was discussed above in Sec.~\ref{sec:Counting}.

\subsection{Ground state preparation}\label{sec:GroundStatePrep}

Having constructed our Hamiltonian with an explicit symmetry group  $G = A_4 \times \parity$, the next step is to obtain the ground states. For this, we use the numerical ``iTEBD" algorithm \cite{iTEBD, TEBD, CanonicalForm} to compute the ground states over a range of parameters, $\lambda \in [-2, 2]$ and $\mu \in [-3, 4]$ (this range is simply chosen based on our results to include a large but not necessarily comprehensive sample of different SPT phases). The algorithm computes the ground state of a Hamiltonian $H$ through the imaginary time evolution of an arbitrary initial state $| \psi \rangle$, since $\ket{\psi}$ can be expanded in the  energy eigenbasis of Hamiltonian as  $| \psi \rangle  = \sum_i c_i | E_i \rangle$ and hence $e^{- \tau H} | \psi \rangle$ will suppress all such components except for the ground state $| E_0 \rangle$ in the large-$\tau$ limit. Except where otherwise noted, data in this paper were prepared with a random initial state represented as an MPS with bond dimension $\chi=24$, and evolved according to a fixed sequence of timesteps which were chosen to be sufficient to converge the energy to the level of $10^{-8}$ at the most numerically ``difficult" states. Within each phase, a random set of points have also been recomputed using states with a series of larger bond dimensions ($\chi = 36, 42$, and $60$) and a longer sequence of imaginary timesteps, in order to verify that the observed characteristics are not likely to be artifacts of the numerical parameters.

While the numerical details of the iTEBD algorithm have been extensively documented elsewhere and are outside the scope of our concern here, there is one salient point which must be remarked upon. For a Hamiltonian $H$ with two-body interactions, the algorithm relies on a decomposition of the Hamiltonian into two sets of terms, those acting first on an even site ($H_A$) and those acting first on an odd site ($H_B$), so that $H = H_A + H_B$. As such the imaginary time evolution operator can be approximated by the Suzuki-Trotter decomposition \cite{SuzukiOriginal, SuzukiTrotter2}, which, to second order, gives 
\begin{equation}\label{eq:iTEBDTrotter}
e^{- \tau H} \approx (e^{-\delta \tau H_A /2}e^{-\delta \tau H_B}e^{-\delta \tau H_A /2}) ^ N,
\end{equation}
with $\delta \tau = \tau/N$. The total operator can then be applied as a sequence of smaller operators, acting either on an even site first, or an odd site first. This distinction, then, requires the state to be represented with at least two tensors, $A^j_A$ and $A^j_B$, even if the the resulting state is expected to possess a one-site translation invariance (which would generally allow it to be represented by only a single tensor $A^j$. This fact will have relevance in later sections, when the translational invariance of the MPS is explicitly discussed). 

For now, however, let us simplify the discussion by considering a translationally invariant, infinite ground state, represented by the tensor $A^j$. Note that there is some gauge freedom allowed in the representation of an infinite MPS state--the tensors $A^j$ and $e^{i \phi}XAX^{-1}$ both represent the same one-site translationally invariant state, for example.  This freedom allows us to make some choices about the structure of the representation which will prove useful in subsequent calculations. In particular, we can choose our MPS to be represented in the so-called ``canonical form" \cite{MPS_Perez-Garcia2007, CanonicalForm}, in which the state satisfies the property,

\begin{equation}\label{eq:CanonicalForm}
A_{\alpha, \beta}^j (A^*)_{\alpha', \beta'}^j \delta^{\beta, \beta'} = \delta_{\alpha, \alpha'}
\end{equation}

\begin{figure}[ht]
\includegraphics[width = 90mm]{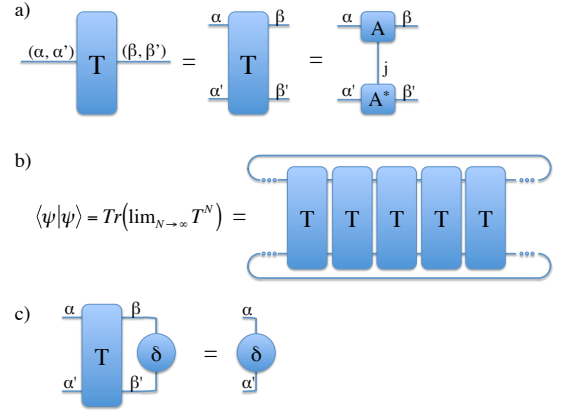}
\caption{\label{fig:TransferMatrix} (Color Online). The transfer matrix of a translationally-invariant matrix product state, demonstrated in graphical tensor notation. In (a), the construction of the transfer matrix is shown as a contraction of two MPS matrices, with the virtual indices grouped to form a single matrix. In (b), the relationship between the transfer matrix and the norm square of the state is shown. Finally, in (c) we show graphically the behavior of a matrix product state in canonical form: such a state has a transfer matrix whose dominant eigenvector is a vectorized version of the identity matrix.}
\end{figure}

This condition can also be thought of in terms of the state's $\emph{transfer matrix}$ (see Fig.~\ref{fig:TransferMatrix}). This object, a common construction used in MPS formalism to compute things like expectation values, is given by:

\begin{equation}\label{eq:Transfer}
T^{(\alpha \alpha')}_{(\beta \beta')} \equiv  A^j_{\alpha, \beta} (A^*)^{j}_{\alpha', \beta'}.
\end{equation}

Now consider the dominant eigenvector of $T$, which will be some vector $X_{(\beta, \beta')}$. Because the outgoing indices of $T$ are a composite of smaller indices $(\beta, \beta')$, any eigenvector of this matrix can also be thought of as a (smaller) matrix in its own right, by interpreting $X_{(\beta, \beta')}$ as $X^{\beta}_{\beta'}$. The original vector $X_{(\beta, \beta')}$ is called the $\emph{vectorization}$ of the matrix $X^{\beta}_{\beta'}$. Now, the condition for canonical form can be rephrased as the requirement that the dominant eigenvector of the state's transfer matrix is a vectorization of the identity matrix, i.e. 

\begin{equation}\label{eq:CanonicalCondition}
(T)^{(\alpha \alpha')}_{(\beta \beta')} \delta^{(\beta \beta')}  = \delta^{(\alpha \alpha')}
\end{equation}
This property of a transfer matrix in canonical form (graphically depicted in Fig.~\ref{fig:TransferMatrix}) will be quite useful in subsequent calculations.
 
Because it represents a contraction of the physical indices of the tensors $A^j$, the transfer matrix can be thought of as containing the overlap of the state with itself at a single site. In other words, in an $N$-site periodic state with one-site translation invariance, the norm square of the state is given by taking a product of $N$ transfer matrices (one for each site) and then tracing over them.

\begin{equation}
\langle \psi | \psi \rangle = Tr[T^N]
\end{equation}

This fact in turn produces a relationship between the eigenvalues $\lambda_j$ of the transfer matrix, and the norm of the state. Consider for example an infinite-length, translation-invariant state with unique largest eigenvalue $\lambda_1$, whose norm is given by $\lim_{N \to \infty} Tr[T^N] = \sum_j \lambda_j^N \approx \lambda_1^N$. This state is normalized if $|\lambda_1| = 1$. Hence in practice, computing the largest eigenvalue of the transfer matrix gives us a convenient way to ensure normalization. 

A general iMPS computed via iTEBD will not necessarily be in exactly canonical form. However, because this form is ultimately so useful, it is worthwhile to enforce it for the ground state representations at the time of their calculation. In \cite{CanonicalForm}, Orus and Vidal have given an analytical prescription was given for placing an arbitrary iMPS in canonical form. However, successive Schmidt decompositions of the state during an iTEBD algorithm are themselves equivalent to enforcing canonical form, so long as the operators being applied to the state are unitary. Of course, when one computes a ground state using imaginary time evolution, the operators which are used, of the form $e^{- \delta \tau H}$ (see Eq.~\ref{eq:iTEBDTrotter}), are not in general unitary. But for $\delta \tau$ very small, they will be quite close. Since a typical iTEBD algorithm ends with a sequence of very small time step evolutions, the resulting states are also typically ``close" to canonical form~\cite{SchachenmayerThesis}. To take this to its logical extension, it is a good practice to terminate every iTEBD algorithm with e.g. $100$ steps of evolution in which we apply only the identity gate (which corresponds to the exact  $\delta \tau = 0$ limit). Of course, this identity gate evolution is both explicitly unitary and incapable of changing the underlying state. In this way, one can ensure that the states computed via iTEBD algorithm are exactly in canonical form (up to numerical precision).

\subsection{Symmetry detection and extraction of order parameters}

\subsubsection{States with one-site translation invariant representations}
The general numerical scheme for extracting the topological order parameters from a numerical MPS was presented in \cite{top1d_pollmanturner}, where it was principally used to study the order parameters $\omega, \beta_{P}$, and $\beta_{T}$, a parameter for time-reversal symmetry. Here we emphasize that it can be used to extract other parameters like the 1D representation $\chi$ as well. We consider the situation first for onsite symmetries and assume that the infinite state possesses one-site translation invariance and is represented by a tensors $A_j$. The generalization to other symmetries and to different levels of translation invariance will be considered subsequently.

\begin{figure}[ht]
\includegraphics[width = 90mm]{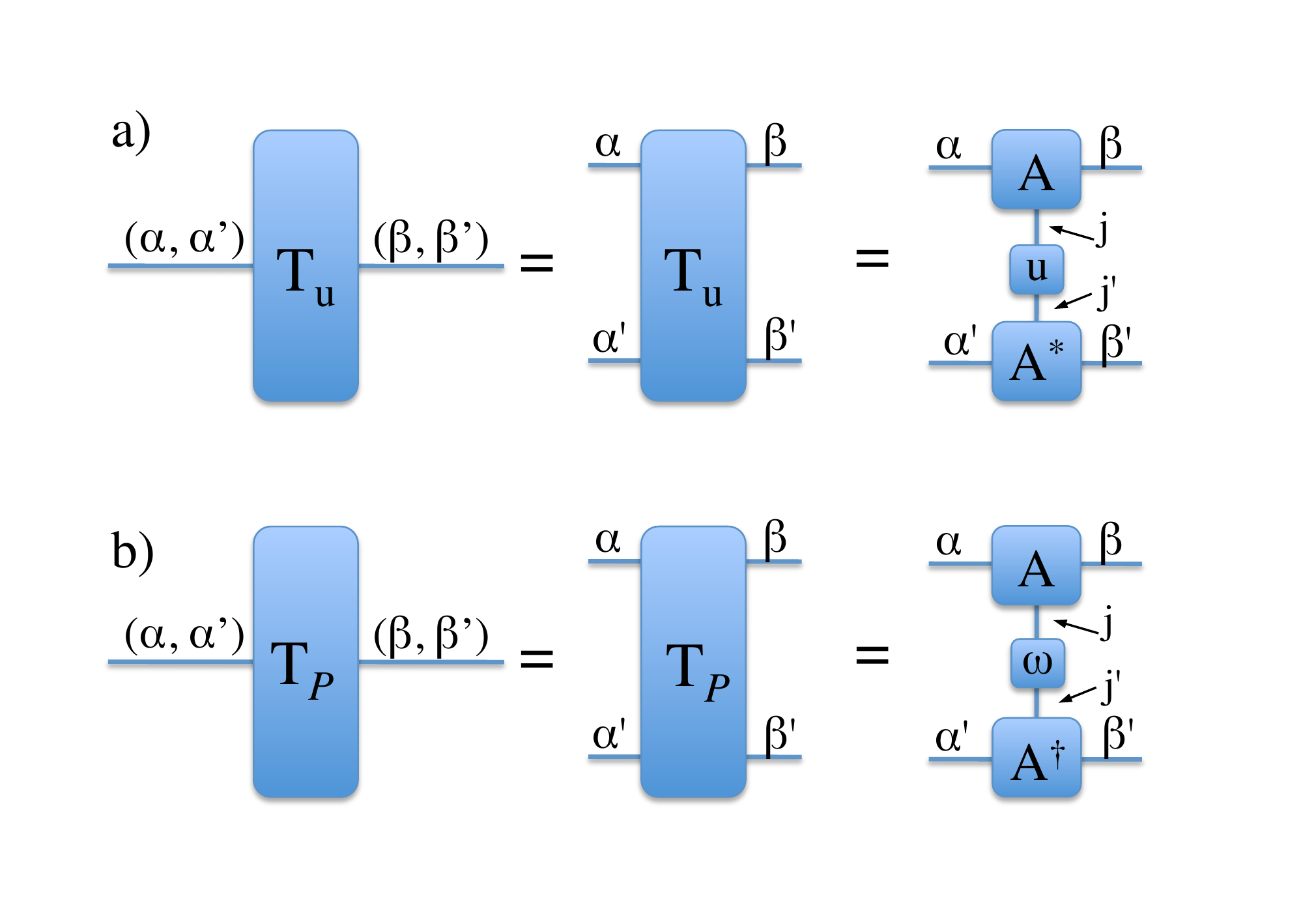}
\caption{\label{fig:GenTransfer} (Color Online). The notion of the transfer matrix can be generalized to include (a) onsite operation $U = \bigotimes_j u_j$, or (b) a parity operation $\parity_\omega$. Generalizations to other symmetries are possible, but outside the scope of this paper as they are not present in our model.}
\end{figure}

To check for symmetry and ultimately access the topological parameters, one first defines a ``generalized" transfer matrix $T_u$, which extends the definition in Eq.~\ref{eq:Transfer} to include the action of some onsite operator $u$ between the physical indices, i.e. 

\begin{equation}\label{eq:GenTransfer}
T_u \equiv A^{j} u_{j, j'} (A^*)^{j'},
\end{equation}
where, this time, we have suppressed the external indices of the matrix (See Fig.~\ref{fig:GenTransfer} for a graphical depiction). In the same manner that the original transfer matrix $T$ represents the contribution of one site to the overlap $\langle \psi | \psi \rangle$, in this case the generalized transfer matrix $T_u$ represents the contribution of one site to the expectation value $\langle \psi | U | \psi \rangle$, where $U = \bigotimes_j u_j$ represents the application of $u$ to every site on the chain (see Fig.~\ref{fig:GenTransfer}). And  just as an iMPS is not normalized unless $T$ has largest eigenvalue $1$, so too is such state only symmetric under $U$ if $T_u$ has largest eigenvalue with unit modulus. 

To study the SPT classification of a state, we thus begin by determining the symmetry. To check if the state is symmetric under the application of $U$, then we first construct $T_u$ and compute the dominant eigenvector $X$ and the associated eigenvalue $\lambda_1$. Note that, when the dimensions of $T_u$ is large, it is numerically far easier to use some iterative procedure such as a power or Lanczos algorithm \cite{Lanczos1950, ImplicitLanczos, EigBook} to extract this, since only the largest eigenvalue is required and not the entire spectrum. If $|\lambda_1| < 1$, the state is not symmetric under $U$ because $\langle \psi | U| \psi \rangle = lim_{N \to \infty} T(u)^N $ will vanish. If, however, the unique largest eigenvalue gives $| \lambda_1| = 1$, then we can proceed with the analysis.

Consider now a normalized iMPS in canonical form, which is invariant under a set of symmetries $u(g)$ at each site for $g$ in some symmetry group $H \in G$. Per Eq.~\ref{eq:mps_invariance} above, this invariance implies the existence of a set of matrices $V(g)$, which are generally projective representations, and $\chi(g)$, a one-dimensional representation. As shown in \cite{top1d_pollmanturner}, one can extract both the projective and a 1-dimensional representation parameters directly from the dominant eigenvector and eigenvalue of the generalized transfer matrix. In particular, if $X$ is the dominant eigenvector (or more precisely, if $X^{\beta}_{\beta'}$ is a matrix and it's vectorization $X_{(\beta \beta')}$ is the dominant eigenvector), then $V = X^{-1}$. The one-dimensional rep $\chi(g)$ is simply equal to the dominant eigenvalue itself. In other words, 

\begin{equation}\label{eq:TransferEigenEquation} 
(T_u)^{(\alpha \alpha')}_{(\beta \beta')} (V^{-1})^{(\beta \beta')}  = \chi \cdot (V^{-1})^{(\alpha \alpha')}
\end{equation}

\begin{figure}[ht]
\includegraphics[width = 90mm]{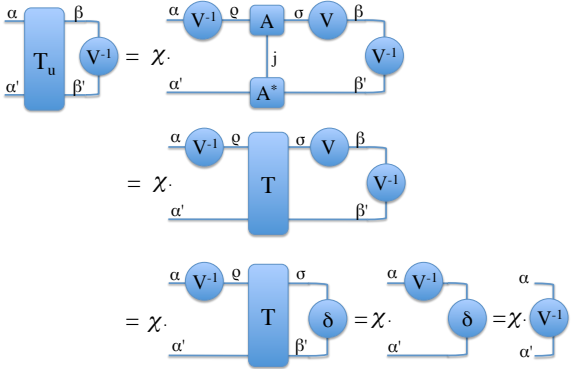}
\label{fig:ParameterExtraction}
\caption{(Color online).The projective representation $V$ of a symmetry can be obtained from a state's generalized transfer matrix because the dominant eigenvector of said matrix will be the vectorization of, $V^{-1}$, so long as the original state is in canonical form. This relation is demonstrated graphically for the case of an onsite symmetry, but easily generalizes to the parity case. }
\end{figure}

To see this, consider the left hand side of the equation (In many ways, this line of argument is clarified when represented by graphical notation; see also Fig.~\ref{fig:ParameterExtraction}). Combining the definition of the generalized transfer matrix, Eq.~\ref{eq:GenTransfer}, with the symmetry fractionalization condition in Eq.~\ref{eq:mps_invariance}, we have 

\begin{align*} 
(T_u)^{(\alpha \alpha')}_{(\beta \beta')} &=  \chi \cdot (V^{-1})^{\alpha \rho} A^j_{\rho \sigma} V^{\sigma \beta} (A^{\dagger})^j_{\alpha' \beta'} \\ 
&= \chi \cdot (V^{-1})^{\alpha \rho} T^{(\rho \alpha')}_{(\sigma \beta')} V^{\sigma \beta}.
\end{align*}

When this is inserted in the left hand side of Eq.~\ref{eq:TransferEigenEquation}, the resulting cancellation of $V$ and $V^{-1}$  gives us

\begin{equation}
(T_u)^{(\alpha \alpha')}_{(\beta \beta')} (V^{-1})_{(\beta \beta')}  =\chi \cdot (V^{-1})^{\alpha \rho} T^{(\rho \alpha')}_{(\sigma \beta')} \delta^{\sigma \beta'}.
\end{equation}
Then, relabeling the dummy indices $\rho$ and $\sigma$ into $\alpha$ and $\beta$, we can appeal to the canonical form condition of Eq.~\ref{eq:CanonicalCondition} to see that 

\begin{equation}
(T_u)^{(\alpha \alpha')}_{(\beta \beta')} (V^{-1})^{(\beta \beta')}  =\chi \cdot (V^{-1})^{\alpha \alpha'},
\end{equation}
which proves that $V^{-1}$ (vectorized) is an eigenvector with eigenvalue $\chi$. Furthermore, because the state is normalized and because we required as a condition for symmetry that$|\chi| = 1$, this proves that $V^{-1}$ is the dominant eigenvector, up to an overall phase factor in $V$. Hence, any procedure to numerically extract the dominant eigenvector and largest eigenvalue from the generalized transfer matrix is sufficient to extract both the 1D representation $\chi$ and the projective representation required to compute the projective parameters $\omega$ as defined above. 

In the foregoing, we have considered only onsite symmetries applied globally to every site on the state\cite{top1d_pollmanturner}. To include other types of symmetries, one simply generalizes further the notion of the already-generalized transfer matrix. For example, the parity symmetry defined by Eq.~\ref{eq:parity_generator} can be studied by means of the matrix

\begin{equation}\label{eq:ParityTransfer}
T_{\parity} \equiv A^{j} w_{j, j'} (A^{\dagger})^{j'}.
\end{equation}

In comparison to Eq.~\ref{eq:GenTransfer}, we have simply inserted the action of the inversion operator $\inversion$ by performing a transpose on the virtual indices of the second MPS tensor. In this way, the resulting generalized transfer matrix still represents a one-site portion of the overlap $\langle \psi | \parity | \psi \rangle$. By the same arguments as above, and by analogy between Eq.~\ref{eq:mps_invariance} and Eq.~\ref{eq:mps_parity}, one can see that the quantities $\alpha(P)$ and $N$ can be extracted from the dominant eigenvector and eigenvalue as before, with the latter used to compute the parity parameter $\beta(P)$ as described above. 

\subsubsection{States without one-site translationally invariant representations}\label{sec:OneSite}

Thus far, we have also assumed a state with one-site translation invariance. However, even when the ground state being studied $\emph{does}$ possess a one-site translational symmetry, the tensors in the MPS $\emph{representation}$ of this state may not, because the gauge freedom of an MPS is not itself constrained to be translationally invariant. For example, consider a set of translationally-invariant tensors $\{ A^{1}, A^{2}, A^{3} \dots \}$ and the gauge transformation
\begin{equation}
A^j \to  
\begin{cases} 
     XA^jZ^{-1}, & j~\text{even} \\
~&~\\
     ZA^jX^{-1}& j~\text{odd}
   \end{cases}  
\end{equation}
for any appropriately-dimensioned matrices X and Z. Such a gauge transformation results in an MPS representation of the state whose tensors at even and odd sites may look dramatically different. But both sets of tensors (before and after the transformation) collectively represent the same, translationally invariant state. Cases like this are of particular interest here because, as noted above, the iTEBD method (like other MPS ground-state preparation algorithms) necessarily results in an MPS representation with different tensors at even and odd sites, regardless of the translational symmetry of the physical state.

This feature does not affect our numerical calculation of the SPT order parameters $\omega$ and $\beta(P)$, which are obtained as eigenvectors of the generalized transfer matrices, but has important significance for the one-dimensional parameters $\chi, \alpha(P)$, and $\gamma(P)$. Consider, for example, a state which is represented by $k$ sets of tensors $ \{ A^{j_1}... A^{j_k} \}$, either because the underlying state has only a $k$-site symmetry, or perhaps simply because our particular numerical representation requires it. The symmetry condition of Eq.~\ref{eq:mps_invariance} must still hold on a $k$-site level; that is, we will have 

\begin{equation}
\label{eq:mps_invariance_N}
u(g)^I_J A^J = \chi(g)^k V^{-1}(g) A^J V(g),
\end{equation}

where $A^J = A^{j_1}A^{j_2} \cdots A^{j_k}$ is now a tensor representing the entire block of spins which are the unit cell of the translation invariance, and the composite indices $I$ and $J$ are equal to $(i_1 i_2 \cdots i_k)$ and $(j_1 j_2 \cdots j_k)$. Clearly if we now define a $k$-site generalized transfer matrix, 

\begin{equation}\label{eq:GenTransferN}
T^{(k)}_u \equiv A^{J} u_{I, J'} (A^*)^{J'}
\end{equation}
then the arguments from the preceding section show that $V^{-1}$ can still be found as the dominant eigenvector of $T^{(k)}_u$. 

The largest eigenvalue, on the other hand, is now equal not to $\chi$, but to $\chi^k$. In the typical case of an iTEBD state, where $k = 2$, this is problematic because for many common symmetry groups, the values of $\chi(g)$ will be $\pm 1$, so a numerical calculation which gives only $\chi^2$ will be unable to distinguish between the different phases. More generally of course, a $k$-site representation will always leave us initially unable to distinguish the cases where $\chi$ is a $k^{th}$ root of unity. 

Of course, if the underlying state has a one-site translation invariance (despite being represented by tensors with only a two-site invariance), one expects that by use of some suitable gauge transformations it should be possible to transform the representation itself back into a translationally-invariant form. Here, we show how this can be done in practice. Suppose we have a translationally-invariant state with, say, a two-site representation $\{A^{j}, B^{j+1}\}$ and an even number of total spins, such that the state in question is given by either
 \begin{eqnarray}\label{eq:TwoSiteState} 
 \ket{\psi} = \sum_{j_1 \dots} Tr[ A^{j_1} B^{j_2} A^{j_3}  \dots B^{j_N}] \ket{j_1 j_2 \ldots j_N} \\
 \text{or}~  \ket{\psi} = \sum_{j_1 \dots} Tr[ B^{j_1} A^{j_2} B^{j_3}  \dots A^{j_N}] \ket{j_1 j_2 \ldots j_N} 
 \end{eqnarray}

To recover a one-site representation, we first construct a new tensor of the form:

\begin{equation}\label{BlockOffDiag}
\tilde{A^j} = \left( \begin{array}{cc}
\boldsymbol{0} & B^{j} \\
A^{j} & \boldsymbol{0}  \end{array} \right). 
\end{equation}

This new tensor in fact describes the same wavefunction $\ket{\psi}$. This can be seen by considering the product:

%This new tensor can itself be thought of as the tensor specifying a one-site MPS, with coefficients given by $Tr[\tilde{A_{J_1}} \tilde{A_{J_2}} \dots]$. But the tensor now contains two equivalent representations of the underlying state, as can be seen by considering the resulting products when the new tensors are contracted.

\begin{equation}\label{eq:TildeProd}
\prod_j \tilde{A_j} = 
       \left( \begin{array}{cc}
  A^{j_1}B^{j_2}A^{j_3}B^{j_4}\cdots & \boldsymbol{0} \\
\boldsymbol{0} & B^{j_1}A^{j_2}B^{j_3}A^{j_4}\cdots   \end{array} \right).
\end{equation}

If we take $\tilde{A^j}$ to be the tensor specifying a new MPS and compute the coefficients, we will have

\begin{equation}
\begin{split}
\ket{\tilde{\psi}} &= \sum_{j_1 \cdots} Tr[\tilde{A}^{j_1} \dots \tilde{A}^{j_N}] \ket{j_1 \dots j_N} \\
&= \sum_{j_1 \cdots} Tr \left(\prod_j \tilde{A_j} \right) \ket{j_1 \dots j_N}\\
\end{split}
\end{equation}
and thus, upon substituting Eq.~\ref{eq:TildeProd}, we find

\begin{align}
\ket{\tilde{\psi}} &= \sum_{j_1 \cdots} Tr[A^{j_1}B^{j_2}\cdots]  \ket{j_1 \dots}\\
&\phantom{\sum_{j_1 \cdots} Tr[A}+ \sum_{j_1 \cdots} Tr[B^{j_1}A^{j_2}\cdots]  \ket{j_1 \dots}\\
 &= 2 \ket{\psi}
\end{align}
 
%\begin{multline}
%\sum_{j_1 \cdots} Tr[\tilde{A}^{j_1} \dots \tilde{A}^{j_N}] \ket{j_1 \dots j_N} = \\
%\sum_{j_1 \cdots} Tr\left( \begin{array}{cc}
%A^{j_1}B^{j_2}A^{j_3}B^{j_4}\cdots & \boldsymbol{0} \\
%\boldsymbol{0} & B^{j_1}A^{j_2}B^{j_3}A^{j_4}\cdots   \end{array} \right) \ket{j_1 \dots j_N}\\
%= \sum_{j_1 \cdots} Tr[A^{j_1}B^{j_2}\cdots]  \ket{j_1 \dots} + \sum_{j_1 \cdots} Tr[B^{j_1}A^{j_2}\cdots]  \ket{j_1 \dots}\\= 2 \ket{\psi}
%\end{multline}

In other words, the state described by the tensor $\tilde{A^j}$ is essentially identical to the state specified by the original tensors $\{ A^j, B^{j+1} \}$. The only difference is that the correct product of tensors needed to give the coefficients of the state in Eq.~\ref{eq:TwoSiteState} will always appear twice, differing only by an irrelevant one-site translation (because the underlying state has a one-site translation invariance to begin with, these two copies of the state are still equivalent). 

Because the new tensor $\tilde{A^j}$ now contains two degenerate descriptions of the same state, it can be placed in a block diagonal form by appealing to the procedure given in Ref. \cite{MPS_Perez-Garcia2007} for block-diagonalizing an MPS representation (see also Appendix C in \cite{CZX}). The resulting blocks will each independently represent the state, but with one-site translation invariance.  

The procedure, briefly outlined, is as follows: first, one must ensure that the tensor $\tilde{A}^j$ is itself in the canonical form, in the sense that it satisfies Eq.~\ref{eq:CanonicalForm}.  To do this, construct the transfer matrix for $\tilde{A}^j$ and compute the dominant eigenvector. This may result in a degenerate manifold of eigenvectors, but by properties of the transfer matrix, at least one of these will be the vectorization of some positive matrix $X$ \cite{evans1978spectral}. Since this $X$ is invertible, we can then take $\tilde{A}^j \to X^{-1/2} \tilde{A}^j X^{1/2}$. By construction this new definition of $\tilde{A}^j$ will satisfy the canonical form.  

From this, we once again construct a transfer matrix and compute its dominant eigenvector(s). At least one corresponds to a matrix $Z$ which is not proportional to the identity matrix (up to numerical precision). Furthermore, since the vectorization of $Z^{\dagger}$ is also an eigenvector of the transfer matrix in canonical form, we can take $Z \to (Z + Z^{\dagger})/2$ so that $Z$ is Hermitian (unless $(Z + Z^{\dagger})/2$ is itself proportional to the identity, in which case one can always choose instead $Z \to i(Z - Z^{\dagger})/2$.) Finally, we compute the largest magnitude eigenvalue $z_1$ of this new matrix $Z$, so that we can construct a matrix $W = \mathbb{1} - (1/z_1) Z$ to be a matrix which is manifestly not full rank. Let $P$ be a projector onto the support of $W$, and $P^{\perp}$ the projector onto its complement. We can now decompose $\tilde{A^j}$ around theses spaces, as 

\begin{equation}\label{eq:ADecomposition}
\tilde{A^j} = P\tilde{A^j}P + P^{\perp}\tilde{A^j}P^{\perp}  + P \tilde{A^j} P^{\perp} + P^{\perp} \tilde{A^j} P. 
\end{equation}

The reason for the construction of the matrix $W$ from a fixed point now becomes clear, as it has been shown that for such matrix $W$ and its associated projector $P$, we have $\tilde{A}^{j}P = P\tilde{A}^{j}P$ \cite{MPS_Perez-Garcia2007}. Consequently, the final term in Eq.~ \ref{eq:ADecomposition}, which represents one of two off-diagonal blocks in $\tilde{A^j}$, vanishes identically. This, in turn, ensures that the remaining off-diagonal block cannot mix with either of the diagonal blocks in any product  $\tilde{A^{j_i}} \tilde{A^{j_{i+1}}} \cdots $. It therefore does not participate in the calculation of the coefficients of the corresponding states, and can be ignored. 

The remaining terms, $ P\tilde{A^j}P$ and  $P^{\perp}\tilde{A^j}P^{\perp}$, represent the relevant blocks along the diagonal of the tensor. We remark that in principle, one may need to carry out the above procedure iteratively for each such block ($ P\tilde{A^j}P$ and  $P^{\perp}\tilde{A^j}P^{\perp}$) to see if further block reduction is possible. But in practice, for the two-site iTEBD ansatz, a single iteration should suffice. Then, by construction of $\tilde{A^j}$, each will be an equivalent representation of the same state, and each can represent the state with only a one-site translation invariance. In other words, if we simply treat $ P\tilde{A^j}P$ as the tensor representing the state, we can use all the procedures in the preceding section to directly compute the entire family of SPT parameters.

An alternative method for extracting the one-dimensional parameters when their values are $k^{th}$ roots of unity would be to compute the ground state with a version of the iTEBD algorithm designed to act on an $n$-site unit cell, where $n$ does not divide $k$. In this case, the dominant eigenvalue of the generalized transfer matrix will be $\chi^n$, from which $\chi$ can now be calculated without ambiguity. Such generalized iTEBD algorithms have been employed successfully (see for example \cite{NiTEBD1}), but may be less numerically stable, and cannot be used for a general state unless one is sure that $n$ is commensurate with the underlying translation invariance of the state. Nevertheless, both methods are possible in practice, and we have used both to cross-check one another in the results presented in this paper.

\subsubsection{States with broken translation invariance}\label{sec:TIBreaking}

Finally, it may also be the case that a state lacks a one-site translationally invariant representation precisely because the ground state is not one-site translationally invariant. When this occurs, one can still compute topological order parameters for onsite symmetries, but only once they and the associated symmetries have been suitably redefined to be consistent with the translational invariance. In other words, if the state has a $k$-site translation invariance and is represented by the $k$ tensors  $\{ A^{j_1} A^{j_2} \dots A^{j_k} \}$, one combines the tensors in the same manner contemplated above, forming a new tensor $A^{J} = A^{j_1} A^{j_2} \cdots A^{j_k}$ with an enlarged physical index which is given by the composite index $J = (j_1 j_2 \dots j_k)$. We then also re-interpret the onsite symmetry operation to be $u^I_J = u^{i_1}_{j_1} \otimes  u^{i_2}_{j_2} \otimes \dots u^{i_k}_{j_k}$ under the same convention. Once again, with the tensors merged so they continue to represent an individual ``unit cell" of the state, then the relation of Eq.~\ref{eq:mps_invariance} will still hold, and we can compute the projective representations of the symmetry from the dominant eigenvalue of the transfer matrix. Unlike the situation described above, however, where the dominant eigenvalue did not give the one-dimensional representation $\chi$ (but rather $\chi^k$), in this case the eigenvalue for the merged cell still gives an order parameter. Indeed, there is no longer a physical meaning to the $k^{th}$ root of the eigenvalue, because one-site translation is no longer a symmetry. 

For such states, it is also essential to carefully verify the level of any residual translation symmetry. As discussed above, the traditional iTEBD algorithm assumes a two-site invariant representation of the state; hence, if this algorithm produces a state which appears to have translation symmetry which is broken on the one-site level but present at a two-site level, it cannot be assumed that two-site translation is a symmetry of the true ground state; such symmetry may instead have been forced by the algorithm. In this work, whenever one-site translation symmetry is broken, we recompute the ground state using a version of iTEBD with a larger (say, four-site) unit cell. If the two-site translation invariance is still present after such a test, it can then be safely assumed to be a genuine property of the true ground state, and not a property forced by the numerical ansatz. In general, an algorithm with an $k$-site ansatz cannot by itself confirm translation invariance at the $k$-site level.

\subsection{Obtaining the SPT labels $\{\omega, \beta(P), \gamma(g)\}$}
It is clear how the one-dimensional representations $\chi$ and $\alpha(P)$ can be used by themselves to label a phase, since each is a single number. Now, however we must discuss how to extract similar numerical labels from the projective representations and other matrices obtained above ($V, N, $ etc). Hence, we must define a procedure to obtain an order parameter from these matrices. A good order parameter that gives us an SPT label has to satisfy the following conditions:
\begin{itemize}
	\item It should be sensitive to the fractionalization of the symmetry at the virtual level.
	\item It should be invariant under the allowed gauge transformations of MPS states $V\mapsto X V X^{-1}$, $V \mapsto e^{i \theta} V$ where $V$ is some symmetry acting on the virtual level. 
\end{itemize}

\subsubsection{$\omega \in \hgc$}
The authors of~\cite{top1d_pollmanturner} show that tracing over products of elements of the form  $V(g_1) V(g_2) \dagr{V}(g_1) \dagr{V}(g_2)$ satisfies both the above requirements and also gives us the information to extract the class of $\omega$. We will now consider $\onsitesymmetrygroup=A_4$ and its subgroups ($H =\ztzt,\ztwo,\zthree$ and the trivial group)  for which $\hgc= \ztwo$ ($H = A_4, \ztzt$) or the trivial group (everything else). For groups which have $\hgc = \ztwo$, we will list order parameters which picks the value $\pm 1$ depending on whether the representation is linear or projective indicating if the SPT phase of matter is trivial or non-trivial. (Note: as defined before, $D$ refers to the bond dimension and $V(g)$ is the representation of onsite symmetry at the virtual level~(\ref{eq:mps_invariance}) )
\begin{enumerate}
	\item \begin{itemize}
		\item $\onsitesymmetrygroup= A_4 = \innerproduct{a,x}{a^3 = x^2 = (ax)^3 = e }$
		\item $\hgcinput{A_4} = \ztwo$
		\item $\omega = \frac{1}{D} Tr\left[(V(a) V(x) V^\dagger(a) V^\dagger(x))^2\right] = \pm 1$
	\end{itemize}
	\item \begin{itemize}
		\item $\onsitesymmetrygroup= \ztzt = \innerproduct{x_1,x_2}{x_1^2 = x_2^2 = (x_1 x_2)^2 = e }$
		\item $\hgcinput{\ztzt} = \ztwo$
		\item $\omega = \frac{1}{D} Tr\left[V(x_1) V(x_2) V^\dagger(x_1) V^\dagger(x_2)\right] = \pm 1$
		\end{itemize}
	\item \begin{itemize}
		\item $\onsitesymmetrygroup= \zthree$ or $\ztwo$ or the trivial group 
		\item $\hgcinput{G} =$ trivial group
		\item $\omega = 1$ (no projective representations)
	\end{itemize}
\end{enumerate}

\subsubsection{$\beta(P)$ and $\gamma(g)$}
It was shown in~\cite{top1d_pollmanturner} that $\beta(P)$ can be obtained as
\begin{equation}
\beta(P) = \frac{1}{D} Tr\left[N N^*\right]
\end{equation}
From Eq~(\ref{eq:onsite+parity_MPS}) we can see that $\gamma(g)$ that results from the commutation of onsite and parity can be obtained as
\begin{equation}\label{eq:GammaParam}
\gamma(g) = \frac{1}{D} Tr\left[N^{-1} V(g) N V^T(g)\right]
\end{equation}

Here, however, an important technical point arises. Although eq~(\ref{eq:GammaParam}) has a similar form to the equations used to compute $\omega$ and $\beta$, it differs in an important respect. Recall that, as calculated above, the matrices $V$ and $N$ are obtained only up to arbitrary overall phase factors. These phases are irrelevant to the calculation of $\omega$ and $\beta$, as both $V$ and $V^*$ appear equally in the equations which define them. In Eq.~\ref{eq:GammaParam}, however, the matrix $V^T$ will fail in general to cancel the phase contributed by $V$.

Since the $V(g)$ can carry a different phase for each $g$, we must find a way to self-consistently fix the phase factors of each. In principle, this can always be done by appealing to the properties of projective representations. The extracted matrices $V$ should satisfy a set of relationships 

\begin{equation}\label{eq:FactorSystem}
V(g_1)V(g_2) = \omega(g_1, g_2) V(g_1 g_2),
\end{equation}
with the phases $\omega(i,j)$ forming the ``factor system" of the representation. Since the matrices which we numerically extract by the above procedure do not automatically satisfy this relationship, let us label them $\tilde{V}$, with $\tilde{V}(g) = \theta_g V(g)$ for some phase factor $\theta_g$. From this, one can conclude that the numerical matrices obey a similar relation:

\begin{equation}
\tilde{V}(g_1)\tilde{V}(g_2) = \frac{\theta_{g_1 g_2}}{\theta_{g_1} \theta_{g_2}} \omega(g_1, g_2) \tilde{V}(g_1 g_2).
\end{equation}

By analogy to Eq.~\ref{eq:FactorSystem}, let us define 

\begin{equation}\label{eq:OmegaTilde}
\tilde{\omega}(g_1, g_2) = \frac{\theta_{g_1 g_2}}{\theta_{g_1} \theta_{g_2}} \omega(g_1, g_2).
\end{equation}
Note that these phases $\tilde{\omega}(g_1, g_2)$ can be computed numerically from $(1/D) Tr [ \tilde{V}(g_1)\tilde{V}(g_2) \tilde{V}(g_1 g_2)^{-1} ]$.  Furthermore, since parity is assumed to be a symmetry of the state in question (if it is not, then the concept of a $\gamma$ parameter is undefined and the phase factors $\theta$ are irrelevant), then we must have $\omega(g_1 g_2)^2 = 1$ \cite{top1d_wennew}. Inverting Eq.~\ref{eq:OmegaTilde} and applying this condition tells us that

\begin{equation}
\theta_{g_1}^2 \theta_{g_2}^2 \tilde{\omega}(g_1, g_2)^2 = \theta_{g_1 g_2}^2.
\end{equation} 

Since the $\tilde{\omega}$ are known, this set of equations, which run over all the group elements $g$, are sufficient to solve for the phases $\theta$. In fact, when $V$ is unitary, it is clear from the definition of $\gamma$ in Eq.~\ref{eq:GammaParam} that only $\theta^2$, and not $\theta$ itself, is needed to correct for the spurious phase factors, which further simplifies the system of equations which must be solved. 

In practice, another convenient way to fix these phase factors is by interpreting the projective representations of the group, $\tilde{V}$ as linear representations of the covering group (or at least, one of the covering groups). For example, in the case of $\ztzt$, the quaternion group $\mathbb{Q}_8$ is a covering group. Hence the elements of the projective representation of $\ztzt$, $V(g)$ can have their overall phases fixed so that they obey the structure of this group; in particular, for the representation of the identity element we must have $V(e)^2 = \mathbb{1}$, and for all others, $V(g)^2 = -\mathbb{1}$.

 \section{Summary and future directions}\label{Summary}
 In this paper, we have studied the phase diagram of a quantum spin-1 lattice with an onsite $\af$ symmetry along with invariance under lattice translation and inversion. Using numerical methods, we obtain the ground state of the Hamiltonian for a range of parameters and using appropriate matrix product state order parameters, we study the phase diagram. In the parameter range we study, we detect 8 gapped phases characterized by a combination of symmetry breaking and symmetry fractionalization. In a recent paper~\cite{SPTOTrans3}, the authors study continuous phase transitions between two SPT phases (which do not break symmetry) and determine that the central charge of the conformal field theory (CFT) that describes that system at the phase boundary has a central charge $c \geq 1$. In our phase diagram, we observe that the phase boundaries separating phases \textbf{B} and \textbf{C} and also \textbf{C} and \textbf{D} by continuous phase transitions are characterized by a CFT with $c \approx 1.35$ which is consistent with $c \geq 1$. However, there is a distinction that must be noted. The authors of Ref~\cite{SPTOTrans3} state their result for phase transitions between two distinct SPT phases protected by onsite symmetries \thatis
 when two phases have linear and projective representations in the virtual space. For our case, the phases \textbf{B}, \textbf{C} and \textbf{D} are distinct because of the presence of translation invariance in addition to the internal $\af$ symmetry. Specifically, the ground states belonging to three phases are invariant under $\af$ transformations up to $U(1)$ factors that corresponds to the three 1D representations of $\af$ rather than projective representations. In fact, all three phases have non-trivial projective representations in the virtual space. Furthermore, the authors of Ref~\cite{SPTOTransChen} conjecture that there can exist no continuous phase transitions between non-trivial SPT phases when the internal symmetry is discrete at all length scales. The phase transitions mentioned above appear to be counter examples. However, at the moment we do not know whether the discrete symmetry in our model is enhanced to a continuous one at the transitions between the $A_4$ SPT phases. It seems, however, that the transitions seen in this model are not the result of fine-tuning, as they appear in a finite range of the parameter $\mu$. These observations suggest that it is interesting to study the nature of phase transitions and the physics involved in the phase boundary when the protecting symmetry has both internal and onsite symmetries.   Further analysis in this direction is left for future exploration. 
 
 \section*{Acknowledgements}
 We acknowledge useful discussions with Lukasz Fidkowski, Ching-Yu Huang, Frank Pollmann and Ari Turner. This work was supported in part by the National Science Foundation under Grants No. PHY 1314748 and No. PHY 1333903. 
 
\bibliography{A4PaperV9}{}
\bibliographystyle{unsrt}

\appendix
\section{Constructing the symmetric Hamiltonian} \label{Hamiltonian_construction}
We provide details of the construction of the Hamiltonian in Sec~\ref{Overview}. We remind the reader that the group of onsite symmetries we consider is $\af$, the alternating group of degree four and the group of even permutations on four elements. The order of this group is 12 and can be enumerated with two generators, 
\begin{eqnarray}
\innerproduct{a,x}{a^3 = x^2 = (ax)^3 = e}.
\end{eqnarray}
The onsite representation, $u(g)$ we consider that the spins transform under is the faithful 3D irrep of $\af$ with generators
\begin{eqnarray}
a = \begin{pmatrix}
0 & 1 & 0\\
0 & 0 & 1\\
1 & 0 & 0
\end{pmatrix}, ~ x = \begin{pmatrix}
1 & 0 & 0 \\
0 & -1 & 0\\
0 & 0 & -1
\end{pmatrix}
\end{eqnarray}
We use group invariant polynomials as building blocks to construct Hermitian operators invariant under group action. A group $G$ invariant $n$-variable polynomial $f(x_1,x_2, \dots  x_n)$ is unchanged when the $n$-tuplet of variables $\left(x_1, x_2 \dots x_n \right)$ is transformed under an $n$-dimensional representation of the group $U(g)$.
\begin{eqnarray}
f(x'_1, x'_2 \dots x'_n) &=& f(x_1, x_2 \dots x_n) \\
x'_i &=&  U(g)_{i j} x_j.
\end{eqnarray}
If we have $n$ Hermitian operators $X_{i=1 \dots n}$ that are $n$-dimensional and transform covariantly like the $n$ variables of the polynomial $x_{i=1 \dots n}$, $i.e.$ $U(g) X_i \dagr{U}(g) = U(g)_{i j} X_j$, then we can elevate the group invariant polynomials to group invariant operators as $f(x_1,x_2, \dots  x_n) \rightarrow f(X_i, X_2 \dots X_n)$ carefully taking into account that unlike the numbers $x_i$, the operators $X_i$ do not commute. 

Since we need three-dimensional operators of $\af$, we consider the set of independent three variable polynomials invariant under the action of the 3D irrep of $\af$~\cite{ramond_group}:
\begin{eqnarray}
f_1(x,y,z) &=& x^2 + y^2 + z^2,\\
f_2(x,y,z) &=& x^4 + y^4 + z^4,\\
f_3(x,y,z) &=& xyz.
\end{eqnarray}
We know that the spin operators $S^i$ satisfying $ [S^i, S^j] = i \epsilon_{ijk} S^k $ transform covariantly under any $SO(3)$ rotation, in particular for the finite set of rotations that corresponds to the subgroup $\af \subset SO(3)$. Thus, to find invariant operators for the three-dimensional representation, we need to take the spin operators in the appropriate three-dimensional basis in terms of the Spin-1 states $\ket{J=1, m_z} \cong \ket{m_z} =\{ \ket{\pm1}, \ket{0}\}$ so as to get the irrep defined above.
\begin{multline}
\ket{x} = \frac{1}{\sqrt{2}} (\ket{-1} - \ket{1}),~ \ket{y} = \frac{i}{\sqrt{2}} (\ket{-1} + \ket{1}),~ \ket{z} = \ket{0}, \nonumber
\end{multline}  and elevate the polynomials $f_1, f_2, f_3$ to operators as
\begin{eqnarray}
F_1 &=& S^x_a S^x_b + S^y_a S^y_b + S^z_a S^z_b,\\
F_2 &=& (S^x_a S^x_b)^2 + (S^y_a S^y_b)^2 + (S^z_a S^z_b)^2 ,\\
F_3 &=& S^x_a S^y_b S^z_c + S^z_a S^x_b S^y_c + S^y_a S^z_b S^x_c \nonumber \\
&+& S^y_a S^x_b S^z_c + S^x_a S^z_b S^y_c + S^z_a S^y_b S^x_c, 
\end{eqnarray}
where the indices $a,b,c$ label any other quantum numbers collectively like lattice sites and can be chosen as per convenience, say to make the operators local as we will do next. As a model Hamiltonian, we could use any function of the invariant operators $F_1$, $F_2$ and $F_3$ and ensure that everything is symmetric under the exchange of lattice labels to impose inversion symmetry. 

We start with the Hamiltonian for the Spin-1 Heisenberg antiferromagnet which is constructed using $F_1$ with $\{a,b\}$ chosen to make the interactions nearest neighbor: 
\begin{eqnarray}
H_{Heis} &=& \sum_i  \vec{S}_i\cdot\vec{S}_{i+1}  , \\
\mbox{where}~\vec{S}_i\cdot\vec{S}_{i+1} &\equiv& S^x_i S^x_{i+1} + S^y_i S^y_{i+1} + S^z_i S^z_{i+1}. \nonumber
\end{eqnarray}
We add the two other combinations to the Heisenberg Hamiltonian so as to break the $SO(3)$ symmetry to $\af$ by using $F_2$ and $F_3$ as follows:
\begin{eqnarray}
H_q &=& \sum_i  \vec{S}^2_i\cdot\vec{S}^2_{i+1} \\
\mbox{where,~} \vec{S}^2_i\cdot\vec{S}^2_{i+1} &\equiv& (S^x_i S^x_{i+1})^2 + (S^y_i S^y_{i+1})^2 + (S^z_i S^z_{i+1})^2,\nonumber
\end{eqnarray}
and
\begin{multline}
H_c = \sum_i [ (S^x S^y)_i S^z_{i+1} + (S^z S^x)_i S^y_{i+1} + (S^y S^z)_i S^x_{i+1} \\
+ (S^y S^x)_i S^z_{i+1} + (S^x S^z)_i S^y_{i+1} + (S^z S^y)_i S^x_{i+1} \\
+ S^x_{i} (S^y S^z)_{i+1} + S^z_{i} (S^x S^y)_{i+1} + S^y_{i} (S^z S^x)_{i+1}  \\
+  S^x_{i} (S^z S^y)_{i+1} + S^z_{i} (S^y S^x)_{i+1} + S^y_{i} (S^x S^z)_{i+1}]. 
\end{multline}
The operators are symmetrized so that the Hamiltonian is invariant under inversion as well as lattice translation. With these pieces, we arrive at the total Hamiltonian which is invariant under $A_4 \times \paritygroup$:
\begin{equation}
H = H_{Heis} + \lambda H_c + \mu H_q.
\end{equation}

\section{Review of classification of SPT phases protected by Time reversal symmetry}
\subsection{Without onsite symmetry or parity} \label{app:timerev}
The time reversal symmetry group $\timerevgroup$ is generated by the anti-unitary action $\timerev$ which is a combination of an onsite unitary operator $v$ and complex conjugation, $\conjug$ 
\begin{equation}
\timerev = v_1 \otimes v_2  \cdots \otimes v_N~\conjug
\end{equation}
where, if the basis at each site $\ket{i}$ is real, the action of $\conjug$ is simply 
\begin{eqnarray}
\conjug: c_{i_1 \ldots i_N} &\rightarrow& c^*_{i_1 \ldots i_N}\\
\conjug: Tr[A_1^{i_1} \ldots A_N^{i_N}] &\rightarrow& Tr[(A_1^{i_1})^* \ldots (A_N^{i_N})^*]
\end{eqnarray}
$\timerev^2 = \pm 1$ in general. However, it was shown in Refs.~\cite{top1d_wenold,top1d_wennew} that only the case of $\timerev^2 = 1$ corresponds to gapped phases and we will consider only this case. $\timerevgroup=\{e,\timerev\}$. The action on the MPS matrices is 
\begin{equation}
\timerev: A_M^i \rightarrow v_{ij} (A_M^j)^*
\end{equation}
If $\timerevgroup$ is a symmetry of the Hamiltonian which is not broken by the ground state  $\ket{\psi}$, we have, under the action of $\timerev$,
\begin{equation}\label{eq:wavefuction_timerev}
\timerev \ket{\psi} = \ket{\psi}.
\end{equation}
Note that the possibility of $\alpha(T)$ analogous to $\alpha(P)$ of Sec.~(\ref{sec:parity}) can be eliminated by re-phasing the spin basis (See Refs.~\cite{top1d_wennew,abhishodh1}). The condition Eq.~(\ref{eq:wavefuction_parity}) can also be imposed on the level of the MPS matrices that describe $\ket{\psi}$:
\begin{equation}\label{eq:mps_timerev}
v_{i j } (A_M^{j})^* = M^{-1} A_M^i M,
\end{equation}
Here, $M$ has the property $M^T = \beta(T) M = \pm M$. \emph{$\beta(T) = \pm 1$ labels the two SPT phases protected by $\timerevgroup$}.~\cite{top1d_wennew}

\subsection{With parity} \label{app:timerev+parity}
If the actions of parity and time reversal commute, the 8 SPT phases protected by $\paritygroup \times \timerevgroup$ are labeled by $\{\alpha(P), \beta(P), \beta(T)\}$ as defined before in Secs~(\ref{sec:parity},\ref{app:timerev}).~\cite{top1d_wennew}.

\subsection{With onsite symmetry} \label{app:timerev+onsite}
If the action of the onsite symmetry transformation $U(g)$ commutes with $\timerev$, we have a similar result to Eq.~(\ref{eq:onsite+parity_MPS}).
\begin{equation}
U(g) \timerev \ket{\psi} = \timerev U(g) \ket{\psi},
\end{equation}
this imposes constraints on the matrix $M$ defined as~\cite{top1d_wennew}.
\begin{equation}\label{eq:onsite+timerev_MPS}
M^{-1} V(g) M = \gamma_T(g) V^*(g).
\end{equation}
The different SPT phases protected by $G \times \timerev$ are labeled by $\{\omega,\beta(T),\gamma_T(g)\}$ where, $\omega \in \hgc$ which satisfy $\omega^2=e$, $\gamma_T \in \mathcal{G}/\mathcal{G}^2$ using the same arguments as Sec~(\ref{sec:onsite+parity}). If translation invariance is also a symmetry, the set of 1D representations $\chi(g)$ in Eq~(\ref{eq:mps_invariance}) which satisfy $\chi(g)^2=1$ also label different phases in addition to the ones already mentioned before.~\cite{top1d_wennew}

\subsection{With onsite and parity }
The different SPT phases protected by $G \times \timerev \times \paritygroup$ are labeled by $\{\omega, \chi(g), \alpha(P), \beta(P), \beta(T), \gamma(g), \gamma_T(g)\}$ where, $\omega \in \hgc$ which satisfy $\omega^2=e$, $\gamma(g)$ and $\gamma_T(g) \in \mathcal{G}/\mathcal{G}^2$, $\chi(g)^2=1$ and $\mathcal{G}$ is the set of 1D representations of $\onsitesymmetrygroup$.~\cite{top1d_wennew}

\end{document}